\documentclass[aps,prl,reprint,groupedaddress
]{revtex4-2}
\usepackage[T1]{fontenc}
\usepackage[utf8]{inputenc} 
\usepackage{amsmath}
\usepackage{amsfonts}
\usepackage{graphicx}
\usepackage{amssymb}
\usepackage{textcomp}
\usepackage{bm}
\usepackage{dsfont}
\usepackage{pgfplots}
\usepackage{siunitx}
\DeclareSIUnit\gauss{G}

\usepackage{mathtools}

\DeclarePairedDelimiter\bra{\langle}{\rvert}
\DeclarePairedDelimiter\ket{\lvert}{\rangle}
\DeclarePairedDelimiterX\braket[2]{\langle}{\rangle}{#1\,\delimsize\vert\,\mathopen{}#2}

\DeclareMathAlphabet\mathbfcal{OMS}{cmsy}{b}{n}

\newcommand{\fig}[1]{Fig.\,\ref{#1}}

\newcommand{\omc}{\omega_{\text{c}}}
\newcommand{\bs}{\boldsymbol}
\newcommand{\D}{{\rm d}}
\newcommand{\I}{{\rm i}}
\newcommand{\E}{{\rm e}}

\newcommand{\ebold}{{\boldsymbol e}}
\newcommand{\qbold}{{\boldsymbol q}}
\newcommand{\rbold}{{\boldsymbol r}}
\newcommand{\vbold}{{\boldsymbol v}}
\newcommand{\wbold}{{\boldsymbol w}}
\newcommand{\Abold}{{\boldsymbol A}}
\newcommand{\Bbold}{{\boldsymbol B}}
\DeclareMathOperator{\Real}{Re}

\begin{document}

 \title{
Emergence of a Landau level structure in dark optical lattices
 }
 \author{Sylvain Nascimbene}
 \email{sylvain.nascimbene@lkb.ens.fr} 
 \author{Jean Dalibard}
\email{jean.dalibard@lkb.ens.fr} 
 \affiliation{Laboratoire Kastler Brossel,  Coll\`ege de France, CNRS, ENS-PSL University, Sorbonne Universit\'e, 11 Place Marcelin Berthelot, 75005 Paris, France} 
 \date{\today}

  \begin{abstract}
An optical flux lattice is a set of light beams that couple different internal states of an atom, thereby producing topological energy bands. Here we present a configuration in which the atoms exhibit a dark state, i.e. an internal state that is not coupled to the light. At large light intensity, the low-energy dynamics is restricted to the dark state, leading to an effective continuum model with a Landau-level-like structure. This structure is dramatically different from that of usual topological optical lattices, which lead to discrete models in the tight-binding limit. For well-chosen atomic species, the proposed system is essentially immune to heating due to photon scattering, making it a highly promising  way to emulate the integer or fractional quantum Hall effect.
 \end{abstract}
 
 \maketitle
 
The generation of topological bands for atoms or photons moving in a periodic potential has been recently the subject of many theoretical and experimental studies (for reviews, see e.g. \cite{cooper_topological_2019,ozawa_topological_2019-1}). It opens the way to a simulation of the quantum Hall effect, from the integer to the fractional  regimes when the role of interactions increases. Among the various schemes that have been considered for atoms, optical flux lattices (OFL) involve laser fields inducing a spatially periodic coupling between different internal states of the atom. Suitable laser configurations give rise to topological Bloch bands, with  a periodic effective magnetic flux allowing cold atoms to experience conditions analogous to those found in quantum Hall states. OFL have the advantage of not requiring any time-modulated parameters (which may lead to undesired heating), while providing a large flux density, typically one or two flux quanta per unit cell \cite{cooper_optical_2011}. However, OFL  schemes proposed so far, when they are aimed at  atomic species such as alkali-metal atoms, may induce significant photon scattering, which limits the available time during which the atomic gas remains cold enough to observe topology-induced effects \cite{juzeliunas_flux_2012,cooper_reaching_2013}. 

In this paper, we propose a modified version of the OFL concept that operates with a so-called `dark state', thus reducing considerably the heating problem due to photon scattering. Dark state lattices \cite{hemmerich_trapping_1995,dum_gauge_1996} have recently been identified as a method for achieving sub-wavelength structures, either in 1D \cite{lacki_nanoscale_2016,wang_dark_2018,kubala_optical_2021} or 2D \cite{gvozdiovas_interference-induced_2023,burba_two_2025}. Here we propose a scheme providing an energy spectrum remarkably close to the Landau level structure for a charged particle in a uniform magnetic field. For realistic parameters, we obtain a series of many equidistant, non-overlapping, topological bands, each with a Chern index equal to 1. Being restricted to the dark internal state, the atoms remain weakly coupled to light, giving rise to delocalized orbitals. This behavior contrasts with other realizations of topological bands in optical lattices, which can be described by discrete lattice models such as the Haldane and Harper-Hofstadter models \cite{miyake_realizing_2013,aidelsburger_realization_2013,jotzu_experimental_2014}. As an illustration of the topological character of the bands, we check that an ideal Fermi gas populating the ground band should exhibit an incompressible bulk surrounded by chiral edge states, and weakly interacting bosons should organize themselves into  an ordered vortex lattice.

\begin{figure}[t!]
\includegraphics[scale=0.86,trim=35 8 0 0]{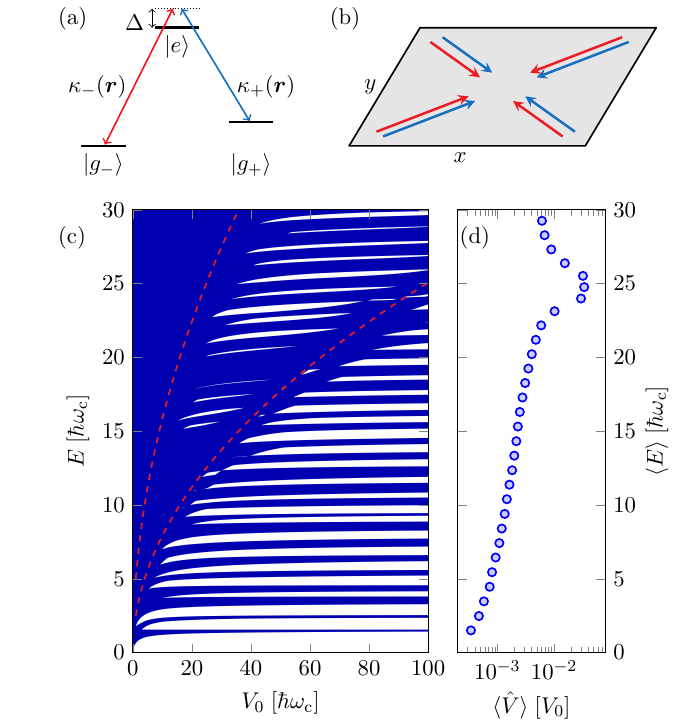}
\caption{(a) The $\Lambda$ level scheme considered in this manuscript, involving light fields at two close frequencies (red and blue arrows). (b) Spatial configuration of the lasers producing the dark optical lattice, with two standing waves along the directions $x\pm y$ for each frequency. (c) Band spectrum of the flux lattice  as a function of the light coupling amplitude $V_0$. In the regime $V_0\gg\hbar\omc$, we find at low energy a succession of narrow energy bands with dominant dark-state character and an almost uniform spacing $\simeq\hbar\omc$. The red lines give a rough estimate of the first two bands with dominant bright-state character. They correspond to the energy levels $(n+1)\hbar\Omega_{\text{bright}}$ (with $n$ integer) in the lattice potential experienced by the bright state in the large depth limit of independent harmonic wells. (d) Mean value $\langle\hat V\rangle$ of the light shift as a function of the mean total energy $\langle E\rangle$, averaged over the quasi-momentum $\qbold$,  for the successive bands. The small value of the light shift for the low-energy bands shows that they predominantly populate the dark state.
\label{fig:bands}}
\end{figure}

\paragraph{The atom-laser coupling.} We consider atoms for which a $\Lambda$-type optical transition can be isolated, with two states $|g_\pm\rangle$ from the ground-level manifold and a state $|e\rangle$ from an  excited manifold (see \fig{fig:bands}a). The OFL is formed by two sets of monochromatic laser beams, with set 1 (resp. 2) at frequency $\omega_1$ (resp. $\omega_2$) close to the frequency $\omega_+$ of the $|g_+\rangle \leftrightarrow |e\rangle$ transition (resp. $\omega_-$ for $|g_-\rangle \leftrightarrow |e\rangle$) (see \fig{fig:bands}b). We assume that these frequencies fulfill the Raman resonance condition $\omega_1-\omega_2=\omega_+-\omega_-$ so that one can identify a space-dependent dark state, linear combination of  $|g_\pm\rangle$ \cite{dum_gauge_1996}.

The atom-laser coupling reads in the $\{|g_+\rangle,|g_-\rangle\}$ basis:
\begin{equation}
\hat V(\bs r)=V_0 \begin{pmatrix}
|\alpha_+|^2 & \alpha_+^*\alpha_- \\ \alpha_+\alpha_-^* & |\alpha_-|^2
\end{pmatrix},
\label{eq:V_DS}
\end{equation}
where $V_0$ is a real, positive energy and $\alpha_\pm(\bs r)$ dimensionless complex amplitudes. Here we adiabatically eliminated the excited state $|e\rangle$ assuming that the detuning from resonance $\Delta\equiv \omega_1-\omega_+=\omega_2-\omega_-$ is positive and large compared to the Rabi frequency characterizing the atom-laser coupling (see appendix 1 for details) \cite{cohen-tannoudji_atom-photon_1998}. 

The dark state $|{\cal D}(\bs r)\rangle \propto \alpha_-(\bs r) |g_+\rangle -\alpha_+(\bs r)|g_-\rangle$ is the eigenstate of $\hat V$ with energy 0. The  other eigenstate of $\hat V$ is the bright state $|{\cal B}(\bs r)\rangle \propto \alpha_+^*(\bs r) |g_+\rangle + \alpha_-^*(\bs r)|g_-\rangle$ with energy $E_{\text{bright}}(\rbold)=V_0(|\alpha_+(\rbold)|^2+|\alpha_-(\rbold)|^2)$. This energy is positive everywhere so that the dark state is always the local ground state of $\hat V(\bs r)$.

In this work, we will choose the following expressions for the coefficients $\alpha_\pm(\bs r)$ \footnote{An earlier version of this manuscript (arXiv:2412.15038v2) used a slightly more involved configuration with
$\alpha_+=1+\frac{1}{2}\left[ \cos(kx)+\cos(ky)\right]$ and $\alpha_-=\sin\left[\frac{k}{2}(x-y)\right]-\I \sin\left[\frac{k}{2}(x+y)\right]$. This configuration was subsequently explored in detail in \cite{burba_two_2025}.}
\begin{eqnarray}
\alpha_+(\bs r)&=&\sin X+{\rm i}\sin Y,
\label{eq:am1}
\\
\alpha_-(\bs r)&=& \cos X+\cos Y,
\label{eq:am2}
\end{eqnarray}
where we set $X=k(x+y)/2$ and $Y=k(y-x)/2$. Here we assumed that the two wavenumbers of the beams forming the lattice, $k_i=\omega_i/c$ ($i=1,2)$, are close to each other and we set $k\sqrt 2\equiv k_1\approx k_2$.  The light-shift coupling $\hat V$ is then invariant upon  discrete translations by $d\,\ebold_x$ and $d\,\ebold_y$   where $d=2\pi/k$, hence a square unit cell of area $d^2$. 

The main results of this paper are based on the numerical calculation of the energy bands and of the associated Bloch states for a two-level atom moving in the potential $V(\bs r)$. This will be done without any assumption regarding the adiabatic following of the dark state. However, it is interesting to start with the case when this adiabatic following is assumed. The adiabatic following of the dark state $|{\mathcal D}(\bs r)\rangle$ gives rise to a synthetic magnetic field (the Berry curvature)  \cite{dalibard_colloquium_2011}, whose flux across the unit cell is quantized by an integer $N_\phi$ times the magnetic flux quantum $h$. The integer $N_\phi$ corresponds to the number of times the  dark state wraps the Bloch sphere associated with the pseudo-spin $|g_\pm\rangle$  when moving across the unit cell \cite{cooper_optical_2011}. Here, one can check that $N_\phi=1$, a non-zero value that characterizes optical flux lattices. For example, the north  pole $|g_+\rangle$ is reached when $\rbold\to \boldsymbol0$, and the south pole $|g_-\rangle$ on the edges of the primitive unit cell $\rbold= \pm (d/2)\ebold_x$ or  $\pm(d/2)\ebold_y$.

This non-zero flux is one of the requirements for an OFL. Another requirement formulated in  \cite{cooper_optical_2011} is that the eigenstates of the coupling $\hat V(\bs r)$ should be non-degenerate at any point $\bs r$. This second requirement is clearly not fulfilled  for the choice (2-3) since $\hat V$ vanishes at the unit cell corner $\rbold\to d(\ebold_x+\ebold_y)/2$. In fact, this second requirement cannot be satisfied for a dark-state based OFL \cite{SuppMat_FL_DS}. As shown in appendix 2, the bright state admixture remains small at any point of the unit cell, so that the local vanishing of $\hat V$ does not prevent the occurence of topological bands.

Note that the configuration studied here yields the maximum possible flux density from light -- about one flux quantum per $\lambda^2$, where $\lambda$ is the wavelength of the beams forming the lattice. This contrasts with early proposals for artificial gauge fields using dark states \cite{juzeliunas_effective_2005,juzeliunas_light-induced_2006} (see also \cite{ruseckas_non-abelian_2005} for a non-Abelian version). Those relied on configurations where $E_{\text{bright}}$ remained non-zero everywhere and assumed adiabatic following of $|{\cal D}(\bs r)\rangle$, producing much lower flux densities -- about one flux quantum per $w^2$ or $\lambda w$, with $w$ the laser beam size.


\paragraph{Band structure.} We show in Fig. \ref{fig:bands}c  the band spectrum for a two-level atom for the Hamiltonian $\hat {\bs p}^2/(2m)+\hat V(\bs r)$ as a function of the light coupling amplitude $V_0$. Here, energies are expressed in units of the cyclotron frequency $\omc=2\pi\hbar/(md^2)$ expected for a Landau level with the same magnetic flux density as our OFL. For large couplings $V_0\gg\hbar\omc$, we find a series of narrow energy bands at low energy, with an almost uniform spacing matching the cyclotron gap $\hbar\omc$. These bands essentially populate the dark state, with a very small residual light shift, shown in \fig{fig:bands}d for $V_0=100\,\hbar\omc$. For this light coupling, the ground band exhibits a width $\simeq0.10\,\hbar\omc$ (about 8 times smaller than the gap to the first excited band), and the mean light shift is as low as $\langle\hat V\rangle\simeq3\times10^{-4}V_0$. The mixing between dark and bright states occurs at a higher energy, which can be estimated by computing the spectrum of an atom subjected to the lattice potential given by the bright-state energy $E_{\text{bright}}(\rbold)$. In the large $V_0$ limit, one can neglect tunneling between the different wells of the bright-state lattice, and one expects a harmonic spectrum for that state of frequency $\Omega_{\text{bright}}=\sqrt{2\pi V_0\omc/\hbar}$ (dashed red lines in \fig{fig:bands}c).

To characterize the topological nature of the bands, we compute their Chern number, and find the value $C=1$ for the first 19 bands at $V_0=100\,\hbar\omc$, confirming the similarity with Landau levels.  Furthermore, the Berry curvature of the ground band is relatively uniform, with an rms deviation equal to $\simeq12\%$ of its mean value.  We study in \cite{SuppMat_FL_DS} the quantum geometrical tensor, which can be used to characterize more precisely  the analogy between the low-lying bands of our OFL and Landau levels. 

A general comparison between OFL bands and ideal Chern bands is provided in \cite{sommer_ideal_2025}, where it is shown that for the lattice described by Eqs.(\ref{eq:am1},\ref{eq:am2}), the width of the ground band tends to 0 as $V_0/\hbar \omega_c\to \infty$. Furthermore, drawing on concepts from moir\'e materials, the authors of \cite{sommer_ideal_2025} demonstrate that, for a fixed $V_0$, this width can be reduced to extremely low values by introducing additional light beams, thereby reinforcing the proximity to the  Landau level structure.


\paragraph{Connection with Landau level orbitals.} We can extend the comparison between the OFL and Landau levels by inspecting individual Bloch states in real space. We first recall the  Landau level formalism in a way that is well suited for comparison with flux lattices, i.e. a spatially periodic problem. We focus here on the ground band, the lowest Landau level (LLL). We assume a magnetic field $\Bbold=B\,\ebold_z$ and adopt the Landau gauge $\Abold=-By\,\ebold_x$, for which the system is invariant upon $x$-translations $t_x(d_x)=\exp(-d_x\,\partial_x)$ of arbitrary distance $d_x$. The resulting conservation of the canonical momentum $p_x$ yields the LLL basis states $\ket{\phi_{p_x}}$,  of wavefunction
\begin{equation}
\phi_{p_x}(x,y) = \frac{\E^{\I p_x x/\hbar}}{\pi^{1/4}\sqrt{\ell}} \exp\left[-\frac{(y - p_x \ell^2/\hbar)^2}{2\ell^2}\right],
\end{equation}
where $\ell$ is the magnetic length. The basis $\ket{\phi_{p_x}}$ is not convenient for comparison with OFL Bloch bands, which are indexed by a 2D quasi-momentum $\qbold$. In order to symmetrize the $x$ and $y$ directions, we use another symmetry of the system, namely the $y$-magnetic translation symmetry \mbox{$t_y(d_y)=\exp[-d_y\,(\partial_y-\I x/\ell^2)]$} \cite{haldane_many-particle_1985}. Translation symmetries obey $t_x(d_x)t_y(d_y)=\E^{-\I d_x d_y/\ell^2}t_y(d_x)t_x(d_y)$, hence a common basis of $x$- and $y$-translations can be found by choosing $d_x=d_y=d=\sqrt{2\pi}\ell$, such that the $d\times d$ unit cell is thread by a single unit of flux quantum $\Phi_0$. Magnetic Bloch states $\ket{\psi_{q_x,q_y}}$ invariant under both $t_x(d)$ and $t_y(d)$ can be formed by discrete sum of momentum states $\ket{\phi_{p_x}}$, as
\begin{equation}
\ket{\psi_{q_x,q_y}}=\sqrt{d}\sum_{n=-\infty}^\infty\E^{\I q_y d n}\ket{\phi_{q_x+nQ}}
\end{equation}
with $Q=2\pi/d$. This expression corresponds to a discrete Fourier transform of the momentum states $\ket{\phi_{q_x+nQ}}$ whose density probabilities are shifted one to the other along $y$ by integer multiples of $d$. The states $\ket{\psi_{q_x,q_y}}$, defined in the Brillouin zone $|q_x|,|q_y|<\pi/d$, form a basis of the LLL. We note that the wavefunctions match the  LLL states defined on a torus of area $d\times d$ thread by magnetic fluxes $2\pi(\Phi_x,\Phi_y)/\Phi_0=(q_x,q_y)d$, given by \cite{haldane_periodic_1985}
\begin{equation*}
\psi_{q_x,q_y}(x,y)=\frac{\E^{-y^2/(2\ell^2)-\I q_x q_y d^2/(2\pi)}}{\pi^{1/4}\sqrt{\ell}}\Theta\begin{bmatrix}q_xd\\ q_yd\end{bmatrix}\left(\frac{x-\I y}{d}\middle|\I\right), 
\end{equation*}
where $\Theta[\begin{subarray}{c}a\\b\end{subarray}](z|\tau)=\sum_{n=-\infty}^\infty\E^{\I\pi\tau(n+a)^2}\E^{\I2\pi(n+a)(z+b)}$ is the Jacobi function \cite{read_quasiholes_1996}.

\begin{figure}[t!]
\includegraphics[scale=0.85,trim=12 9 0 0]{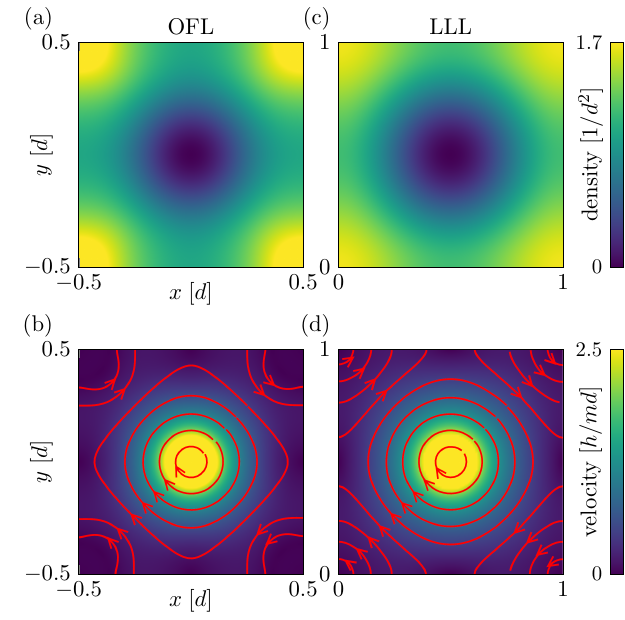}
\caption{Mean density (a,c) and velocity (b,d) distributions on a square unit cell $d\times d$ thread by one unit of magnetic flux quantum.  The left (resp. right) panels correspond to  the lowest band of the OFL for $V_0=100\,\hbar\omc$ (resp. the LLL). In (b,d), the color encodes the norm of the velocity, and the red arrows the velocity streamlines. Similarly to the vortex singularity encountered in the LLL, the unit cell of the OFL exhibits a highly contrasted density dip around which the velocity circulate.
\label{fig:density_velocity}}
\end{figure}


\paragraph{Vortex structure.} We can now compare the properties of magnetic Bloch states for the OFL and LLL. The state $\qbold=\boldsymbol0$ of the OFL exhibits, within each unit cell, a highly contrasted dip of the total density (see \fig{fig:density_velocity}a), around which the velocity circulates (see \fig{fig:density_velocity}b). A similar behavior is found for different quasi-momentum states, albeit with a different hole position. This behavior strongly resembles magnetic Bloch states of the LLL, which exhibit  one quantized vortex point defect per unit cell, at which the density exactly cancels (see \fig{fig:density_velocity}c) and around which the complex phase $\theta$ winds once by $2\pi$. The velocity field, given by
\begin{equation}
\vbold=\frac{1}{m|\psi^2|}\Real\left[\psi^*\left(\frac{\hbar}{\I}\boldsymbol\nabla-q\Abold\right)\psi\right]=\frac{\hbar}{m}\boldsymbol\nabla\theta-\frac{q\Abold}{m},
\end{equation}
is shown in \fig{fig:density_velocity}d. We note that the circulation of velocity around each unit cell cancels due to the compensation between the singular contribution of the vortex and the rotational flow induced by the magnetic field. Each magnetic Bloch state thus exhibits non-zero local currents within the magnetic unit cell, but no global flow on larger length scales.  The occurrence of vortices in OFL orbitals provides another illustration of the `ideal' Chern band character: like in the LLL, OFL orbitals can be fully characterized by the location of vortices  \cite{ledwith_vortexability_2023}.

\begin{figure}[t!]
\includegraphics[scale=0.89,trim=12 8 0 0]{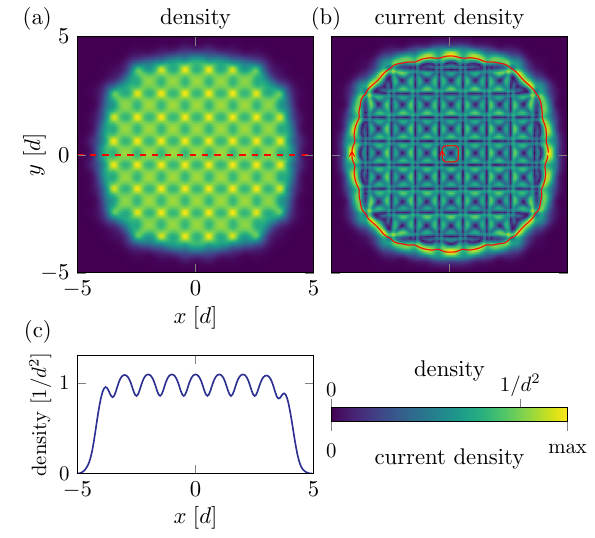}
\caption{In situ density (a) and current density (b) profile of a Fermi gas at zero temperature in a quasi-2D geometry subjected to the optical flux lattice with $V_0=100\,\hbar\omc$ and to an isotropic harmonic trap (trapping frequency $\omega_{\text{trap}}=0.04\,\omc$). We used a chemical potential $\mu_0=1.79\,\hbar\omc$ located between the ground and first excited bands. A few stream lines are shown as red lines in (b). (c) Density profile along the line $y=0$, showing incompressibility in the bulk.
\label{fig:Fermi_sea}}
\end{figure}


\paragraph{Topological signatures with quantum gases.} 
We now consider experimental signatures of the topological character of the low-energy bands of the OFL.  We first consider a low-temperature Fermi gas in a harmonic trap $U(\rbold)$ and subjected to the OFL. We show in \fig{fig:Fermi_sea}a its total density variation, assuming a chemical potential $\mu_0$ set in the gap separating the ground and first excited bands. In the local density approximation framework, we consider the system at position $\rbold$ as locally homogeneous with a chemical potential $\mu(\rbold)=\mu_0-U(\rbold)$. In the central disk defined by $\mu(\rbold)$ higher than the ground band energy, the Fermi sea forms a band insulator, with a uniform coarse-grained density (see \fig{fig:Fermi_sea}c). Outside this disk, the density drops to zero, and the system develops on the edge a chiral current consistent with the bulk-edge correspondence (see \fig{fig:Fermi_sea}b) \cite{halperin_quantized_1982,hatsugai_chern_1993}.

\begin{figure}[t!]
\includegraphics[scale=0.84,trim=11 8 0 0]{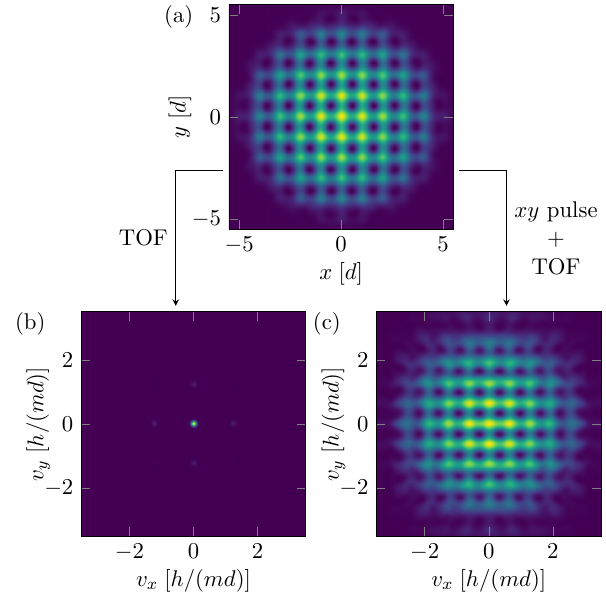}
\caption{(a) In situ density profile of a Bose-Einstein condensate in a quasi-2D geometry subjected to the optical flux lattice with $V_0=50\,\hbar\omc$ and to an isotropic harmonic trap. We used a trapping frequency $\omega_{\text{trap}}=0.06\,\omc$ and an interaction coupling constant $g$ such that $gN/d^2=100\,\hbar\omc$, where $N$ is the number of atoms per spin state. (b) Velocity distribution obtained after a free expansion. The discrete Bragg peaks do not contain information on the vortex lattice structure prior expansion. (c) Velocity distribution obtained after a short pulse of quadrupolar potential $Qxy$ mapping the condensate wavefunction to the symmetric gauge. In that case, velocity distribution reveals the vortex lattice structure.
\label{fig:lattice}}
\end{figure}

We also consider a Bose-Einstein condensate prepared in similar conditions, in the presence of weak repulsive interactions. The condensate wavefunction is computed by evolving the Gross-Pitaevskii equation in imaginary time  using a split-step method. The in situ density profile, shown in \fig{fig:lattice}a, exhibits a square lattice of density holes corresponding to the regular array of quantized vortices expected for the quasi-momentum $\qbold=\boldsymbol0$. We point that, in our calculation, interactions are strong enough to significantly promote atoms to the first excited band (chemical potential $\mu\simeq 1.34\,\hbar\omc$), which leads to a reduction of the  vortex core size compared to the state $\qbold=\boldsymbol0$ of the ground band, shown in \fig{fig:density_velocity}a. 

This vortex lattice shares similarities with those observed in rotating Bose gases \cite{madison_vortex_2000,abo-shaeer_observation_2001,schweikhard_rapidly_2004}, albeit arranged here on a square lattice  imposed by the OFL geometry. In practice, vortices can be revealed as holes in the density profile \cite{mukherjee_crystallization_2022}. One can also consider observing them after a time-of-flight expansion in order to magnify them and reveal the associated quantized phase winding. In our OFL implementation, the velocity distribution cannot be used as such to reveal the vortex lattice, because it only expands over discrete Bragg peaks (see \fig{fig:lattice}b) \cite{baur_adiabatic_2013}. On the other hand, it was shown in  \cite{read_free_2003} that any state of the lowest Landau level in the symmetric gauge expands in time-of-flight in a self-similar manner.  In order to retrieve the vortex lattice configuration in time-of-flight, the condensate wavefunction should be multiplied by the factor $\exp(\I\pi xy/d^2)$ corresponding to the gauge transform from the Landau to symmetric gauge, which can be performed by applying a quadrupolar potential $Qxy$ for a short duration prior to expansion. This is confirmed by calculating the velocity distribution of the condensate wavefunction multiplied by the complex phase factor (see \fig{fig:lattice}c). Note that triangular vortex arrays can also be achieved with dark state lattices \cite{SuppMat_FL_DS,sommer_ideal_2025}.



\paragraph{Implementation.} Different optical setups can be envisioned for implementing the proposed OFL. Here we consider an atom that possesses a narrow transition between a ground level $G$ with angular momentum $J_G$ and an excited level $E$ with angular momentum $J_E=J_G-1$, as found in Lanthanide species such as dysprosium or erbium. We extract a $\Lambda$ system by considering the three magnetic states $|g_-\rangle \equiv |G,m=J_G-1\rangle$, $|g_+\rangle \equiv |G,m=J_G\rangle$ and $|e\rangle\equiv|E,m=J_E\rangle$. The degeneracy between $|g_\pm\rangle$ is lifted using a static magnetic field oriented along the $z$ axis, and the other states from the $G$ and the $E$ manifolds are put out of resonance using the light shift of an auxiliary off-resonant laser beam. All laser beams creating the OFL propagate in the $xy$ plane. Those driving the transition $|g_-\rangle \leftrightarrow |e\rangle$ are linearly polarized along $z$, as needed for a $\Delta m=0$ transition. For the beams driving the transition $|g_+\rangle \leftrightarrow |e\rangle$, we choose an in-plane polarization that can be decomposed as left- and right-handed circular polarization ($\sigma_\pm$) along the $z$-axis. Only the $\sigma_-$ component is relevant for the considered transition. As an exemple we discuss in appendix 3 the practical implementation for Dy atoms and estimate the residual photon scattering rate ($\simeq\SI{1}{\second^{-1}}$) for realistic experimental parameters. 

We also considered the effect of imperfect realization of the atom-laser coupling  and the influence of a residual deviation from the Raman resonance transition on the band structure. As discussed in \cite{SuppMat_FL_DS}, fine tuning of these parameters can be used to flatten the ground band dispersion, and topological features are robust with respect to moderate uncontrolled imperfections.

In conclusion, we have proposed a generalization of optical flux lattices to a dark internal state of an atom. This scheme gives rise to  continuum topological bands even in the presence  of a strong optical lattice. Those bands  can be described as `ideal' Chern bands to a very good approximation, so that they inherit the algebraic structure of Landau levels. As shown in \cite{wang_exact_2021}, this property guarantees the existence of many-body ground states described  by model wavefunctions, such as the bosonic Laughlin state at half filling. This, combined with the strong reduction of spontaneous emission in dark internal states,  makes our proposal appealing for the simulation of quantum Hall effect in atomic gases.

 \begin{acknowledgments}
We acknowledge insightful discussions with  J\'er\^ome Beugnon, Nigel Cooper, Nathan Goldman, Zoran Hadzibabic and Raphael Lopes. We also thank Gediminas Juzeliunas for remarks on an earlier version of this article. This research was funded by  European Union (grant TOPODY 756722 from the European Research Council), Institut Universitaire de France, and Agence Nationale de la Recherche (ANR), projects ANR-24-CE30-7961 and ANR-24-CE47-2670-01. 
\end{acknowledgments}


\begin{thebibliography}{41}%
\makeatletter
\providecommand \@ifxundefined [1]{%
 \@ifx{#1\undefined}
}%
\providecommand \@ifnum [1]{%
 \ifnum #1\expandafter \@firstoftwo
 \else \expandafter \@secondoftwo
 \fi
}%
\providecommand \@ifx [1]{%
 \ifx #1\expandafter \@firstoftwo
 \else \expandafter \@secondoftwo
 \fi
}%
\providecommand \natexlab [1]{#1}%
\providecommand \enquote  [1]{``#1''}%
\providecommand \bibnamefont  [1]{#1}%
\providecommand \bibfnamefont [1]{#1}%
\providecommand \citenamefont [1]{#1}%
\providecommand \href@noop [0]{\@secondoftwo}%
\providecommand \href [0]{\begingroup \@sanitize@url \@href}%
\providecommand \@href[1]{\@@startlink{#1}\@@href}%
\providecommand \@@href[1]{\endgroup#1\@@endlink}%
\providecommand \@sanitize@url [0]{\catcode `\\12\catcode `\$12\catcode
  `\&12\catcode `\#12\catcode `\^12\catcode `\_12\catcode `\%12\relax}%
\providecommand \@@startlink[1]{}%
\providecommand \@@endlink[0]{}%
\providecommand \url  [0]{\begingroup\@sanitize@url \@url }%
\providecommand \@url [1]{\endgroup\@href {#1}{\urlprefix }}%
\providecommand \urlprefix  [0]{URL }%
\providecommand \Eprint [0]{\href }%
\providecommand \doibase [0]{https://doi.org/}%
\providecommand \selectlanguage [0]{\@gobble}%
\providecommand \bibinfo  [0]{\@secondoftwo}%
\providecommand \bibfield  [0]{\@secondoftwo}%
\providecommand \translation [1]{[#1]}%
\providecommand \BibitemOpen [0]{}%
\providecommand \bibitemStop [0]{}%
\providecommand \bibitemNoStop [0]{.\EOS\space}%
\providecommand \EOS [0]{\spacefactor3000\relax}%
\providecommand \BibitemShut  [1]{\csname bibitem#1\endcsname}%
\let\auto@bib@innerbib\@empty
\bibitem [{\citenamefont {Cooper}\ \emph {et~al.}(2019)\citenamefont {Cooper},
  \citenamefont {Dalibard},\ and\ \citenamefont
  {Spielman}}]{cooper_topological_2019}%
  \BibitemOpen
  \bibfield  {author} {\bibinfo {author} {\bibfnamefont {N.~R.}\ \bibnamefont
  {Cooper}}, \bibinfo {author} {\bibfnamefont {J.}~\bibnamefont {Dalibard}},\
  and\ \bibinfo {author} {\bibfnamefont {I.~B.}\ \bibnamefont {Spielman}},\
  }\bibfield  {title} {\bibinfo {title} {Topological bands for ultracold
  atoms},\ }\href@noop {} {\bibfield  {journal} {\bibinfo  {journal} {Rev. Mod.
  Phys.}\ }\textbf {\bibinfo {volume} {91}},\ \bibinfo {pages} {015005}
  (\bibinfo {year} {2019})}\BibitemShut {NoStop}%
\bibitem [{\citenamefont {Ozawa}\ \emph {et~al.}(2019)\citenamefont {Ozawa},
  \citenamefont {Price}, \citenamefont {Amo}, \citenamefont {Goldman},
  \citenamefont {Hafezi}, \citenamefont {Lu}, \citenamefont {Rechtsman},
  \citenamefont {Schuster}, \citenamefont {Simon}, \citenamefont {Zilberberg},\
  and\ \citenamefont {Carusotto}}]{ozawa_topological_2019-1}%
  \BibitemOpen
  \bibfield  {author} {\bibinfo {author} {\bibfnamefont {T.}~\bibnamefont
  {Ozawa}}, \bibinfo {author} {\bibfnamefont {H.~M.}\ \bibnamefont {Price}},
  \bibinfo {author} {\bibfnamefont {A.}~\bibnamefont {Amo}}, \bibinfo {author}
  {\bibfnamefont {N.}~\bibnamefont {Goldman}}, \bibinfo {author} {\bibfnamefont
  {M.}~\bibnamefont {Hafezi}}, \bibinfo {author} {\bibfnamefont
  {L.}~\bibnamefont {Lu}}, \bibinfo {author} {\bibfnamefont {M.~C.}\
  \bibnamefont {Rechtsman}}, \bibinfo {author} {\bibfnamefont {D.}~\bibnamefont
  {Schuster}}, \bibinfo {author} {\bibfnamefont {J.}~\bibnamefont {Simon}},
  \bibinfo {author} {\bibfnamefont {O.}~\bibnamefont {Zilberberg}},\ and\
  \bibinfo {author} {\bibfnamefont {I.}~\bibnamefont {Carusotto}},\ }\bibfield
  {title} {\bibinfo {title} {Topological photonics},\ }\href@noop {} {\bibfield
   {journal} {\bibinfo  {journal} {Rev. Mod. Phys.}\ }\textbf {\bibinfo
  {volume} {91}},\ \bibinfo {pages} {015006} (\bibinfo {year}
  {2019})}\BibitemShut {NoStop}%
\bibitem [{\citenamefont {Cooper}(2011)}]{cooper_optical_2011}%
  \BibitemOpen
  \bibfield  {author} {\bibinfo {author} {\bibfnamefont {N.~R.}\ \bibnamefont
  {Cooper}},\ }\bibfield  {title} {\bibinfo {title} {Optical {{Flux Lattices}}
  for {{Ultracold Atomic Gases}}},\ }\href@noop {} {\bibfield  {journal}
  {\bibinfo  {journal} {Phys. Rev. Lett.}\ }\textbf {\bibinfo {volume} {106}}
  (\bibinfo {year} {2011})}\BibitemShut {NoStop}%
\bibitem [{\citenamefont {Juzeli{\=u}nas}\ and\ \citenamefont
  {Spielman}(2012)}]{juzeliunas_flux_2012}%
  \BibitemOpen
  \bibfield  {author} {\bibinfo {author} {\bibfnamefont {G.}~\bibnamefont
  {Juzeli{\=u}nas}}\ and\ \bibinfo {author} {\bibfnamefont {I.~B.}\
  \bibnamefont {Spielman}},\ }\bibfield  {title} {\bibinfo {title} {Flux
  lattices reformulated},\ }\href@noop {} {\bibfield  {journal} {\bibinfo
  {journal} {New J. Phys.}\ }\textbf {\bibinfo {volume} {14}},\ \bibinfo
  {pages} {123022} (\bibinfo {year} {2012})}\BibitemShut {NoStop}%
\bibitem [{\citenamefont {Cooper}\ and\ \citenamefont
  {Dalibard}(2013)}]{cooper_reaching_2013}%
  \BibitemOpen
  \bibfield  {author} {\bibinfo {author} {\bibfnamefont {N.~R.}\ \bibnamefont
  {Cooper}}\ and\ \bibinfo {author} {\bibfnamefont {J.}~\bibnamefont
  {Dalibard}},\ }\bibfield  {title} {\bibinfo {title} {Reaching {{Fractional
  Quantum Hall States}} with {{Optical Flux Lattices}}},\ }\href@noop {}
  {\bibfield  {journal} {\bibinfo  {journal} {Phys. Rev. Lett.}\ }\textbf
  {\bibinfo {volume} {110}},\ \bibinfo {pages} {185301} (\bibinfo {year}
  {2013})}\BibitemShut {NoStop}%
\bibitem [{\citenamefont {Hemmerich}\ \emph {et~al.}(1995)\citenamefont
  {Hemmerich}, \citenamefont {Weidem{\"u}ller}, \citenamefont {Esslinger},
  \citenamefont {Zimmermann},\ and\ \citenamefont
  {H{\"a}nsch}}]{hemmerich_trapping_1995}%
  \BibitemOpen
  \bibfield  {author} {\bibinfo {author} {\bibfnamefont {A.}~\bibnamefont
  {Hemmerich}}, \bibinfo {author} {\bibfnamefont {M.}~\bibnamefont
  {Weidem{\"u}ller}}, \bibinfo {author} {\bibfnamefont {T.}~\bibnamefont
  {Esslinger}}, \bibinfo {author} {\bibfnamefont {C.}~\bibnamefont
  {Zimmermann}},\ and\ \bibinfo {author} {\bibfnamefont {T.}~\bibnamefont
  {H{\"a}nsch}},\ }\bibfield  {title} {\bibinfo {title} {Trapping {{Atoms}} in
  a {{Dark Optical Lattice}}},\ }\href@noop {} {\bibfield  {journal} {\bibinfo
  {journal} {Phys. Rev. Lett.}\ }\textbf {\bibinfo {volume} {75}},\ \bibinfo
  {pages} {37} (\bibinfo {year} {1995})}\BibitemShut {NoStop}%
\bibitem [{\citenamefont {Dum}\ and\ \citenamefont
  {Olshanii}(1996)}]{dum_gauge_1996}%
  \BibitemOpen
  \bibfield  {author} {\bibinfo {author} {\bibfnamefont {R.}~\bibnamefont
  {Dum}}\ and\ \bibinfo {author} {\bibfnamefont {M.}~\bibnamefont {Olshanii}},\
  }\bibfield  {title} {\bibinfo {title} {Gauge {{Structures}} in {{Atom-Laser
  Interaction}}: {{Bloch Oscillations}} in a {{Dark Lattice}}},\ }\href@noop {}
  {\bibfield  {journal} {\bibinfo  {journal} {Phys. Rev. Lett.}\ }\textbf
  {\bibinfo {volume} {76}},\ \bibinfo {pages} {1788} (\bibinfo {year}
  {1996})}\BibitemShut {NoStop}%
\bibitem [{\citenamefont {{\L}{\k a}cki}\ \emph {et~al.}(2016)\citenamefont
  {{\L}{\k a}cki}, \citenamefont {Baranov}, \citenamefont {Pichler},\ and\
  \citenamefont {Zoller}}]{lacki_nanoscale_2016}%
  \BibitemOpen
  \bibfield  {author} {\bibinfo {author} {\bibfnamefont {M.}~\bibnamefont
  {{\L}{\k a}cki}}, \bibinfo {author} {\bibfnamefont {M.~A.}\ \bibnamefont
  {Baranov}}, \bibinfo {author} {\bibfnamefont {H.}~\bibnamefont {Pichler}},\
  and\ \bibinfo {author} {\bibfnamefont {P.}~\bibnamefont {Zoller}},\
  }\bibfield  {title} {\bibinfo {title} {Nanoscale ``{{Dark State}}'' {{Optical
  Potentials}} for {{Cold Atoms}}},\ }\href@noop {} {\bibfield  {journal}
  {\bibinfo  {journal} {Phys. Rev. Lett.}\ }\textbf {\bibinfo {volume} {117}},\
  \bibinfo {pages} {233001} (\bibinfo {year} {2016})}\BibitemShut {NoStop}%
\bibitem [{\citenamefont {Wang}\ \emph {et~al.}(2018)\citenamefont {Wang},
  \citenamefont {Subhankar}, \citenamefont {Bienias}, \citenamefont {{\L}{\k
  a}cki}, \citenamefont {Tsui}, \citenamefont {Baranov}, \citenamefont
  {Gorshkov}, \citenamefont {Zoller}, \citenamefont {Porto},\ and\
  \citenamefont {Rolston}}]{wang_dark_2018}%
  \BibitemOpen
  \bibfield  {author} {\bibinfo {author} {\bibfnamefont {Y.}~\bibnamefont
  {Wang}}, \bibinfo {author} {\bibfnamefont {S.}~\bibnamefont {Subhankar}},
  \bibinfo {author} {\bibfnamefont {P.}~\bibnamefont {Bienias}}, \bibinfo
  {author} {\bibfnamefont {M.}~\bibnamefont {{\L}{\k a}cki}}, \bibinfo {author}
  {\bibfnamefont {T.-C.}\ \bibnamefont {Tsui}}, \bibinfo {author}
  {\bibfnamefont {M.~A.}\ \bibnamefont {Baranov}}, \bibinfo {author}
  {\bibfnamefont {A.~V.}\ \bibnamefont {Gorshkov}}, \bibinfo {author}
  {\bibfnamefont {P.}~\bibnamefont {Zoller}}, \bibinfo {author} {\bibfnamefont
  {J.~V.}\ \bibnamefont {Porto}},\ and\ \bibinfo {author} {\bibfnamefont
  {S.~L.}\ \bibnamefont {Rolston}},\ }\bibfield  {title} {\bibinfo {title}
  {Dark {{State Optical Lattice}} with a {{Subwavelength Spatial Structure}}},\
  }\href@noop {} {\bibfield  {journal} {\bibinfo  {journal} {Phys. Rev. Lett.}\
  }\textbf {\bibinfo {volume} {120}},\ \bibinfo {pages} {083601} (\bibinfo
  {year} {2018})}\BibitemShut {NoStop}%
\bibitem [{\citenamefont {Kubala}\ \emph {et~al.}(2021)\citenamefont {Kubala},
  \citenamefont {Zakrzewski},\ and\ \citenamefont {{\L}{\k
  a}cki}}]{kubala_optical_2021}%
  \BibitemOpen
  \bibfield  {author} {\bibinfo {author} {\bibfnamefont {P.}~\bibnamefont
  {Kubala}}, \bibinfo {author} {\bibfnamefont {J.}~\bibnamefont {Zakrzewski}},\
  and\ \bibinfo {author} {\bibfnamefont {M.}~\bibnamefont {{\L}{\k a}cki}},\
  }\bibfield  {title} {\bibinfo {title} {Optical lattice for a tripodlike
  atomic level structure},\ }\href@noop {} {\bibfield  {journal} {\bibinfo
  {journal} {Phys. Rev. A}\ }\textbf {\bibinfo {volume} {104}},\ \bibinfo
  {pages} {053312} (\bibinfo {year} {2021})}\BibitemShut {NoStop}%
\bibitem [{\citenamefont {Gvozdiovas}\ \emph {et~al.}(2023)\citenamefont
  {Gvozdiovas}, \citenamefont {Spielman},\ and\ \citenamefont
  {Juzeli{\=u}nas}}]{gvozdiovas_interference-induced_2023}%
  \BibitemOpen
  \bibfield  {author} {\bibinfo {author} {\bibfnamefont {E.}~\bibnamefont
  {Gvozdiovas}}, \bibinfo {author} {\bibfnamefont {I.~B.}\ \bibnamefont
  {Spielman}},\ and\ \bibinfo {author} {\bibfnamefont {G.}~\bibnamefont
  {Juzeli{\=u}nas}},\ }\bibfield  {title} {\bibinfo {title}
  {Interference-induced anisotropy in a two-dimensional dark-state optical
  lattice},\ }\href@noop {} {\bibfield  {journal} {\bibinfo  {journal} {Phys.
  Rev. A}\ }\textbf {\bibinfo {volume} {107}},\ \bibinfo {pages} {033328}
  (\bibinfo {year} {2023})}\BibitemShut {NoStop}%
\bibitem [{\citenamefont {Burba}\ and\ \citenamefont
  {Juzeli{\=u}nas}(2025)}]{burba_two_2025}%
  \BibitemOpen
  \bibfield  {author} {\bibinfo {author} {\bibfnamefont {D.}~\bibnamefont
  {Burba}}\ and\ \bibinfo {author} {\bibfnamefont {G.}~\bibnamefont
  {Juzeli{\=u}nas}},\ }\bibfield  {title} {\bibinfo {title} {Two dimensional
  sub-wavelength topological dark state lattices},\ }\href@noop {} {\bibfield
  {journal} {\bibinfo  {journal} {arXiv:2506.17096}\ } (\bibinfo {year}
  {2025})}\BibitemShut {NoStop}%
\bibitem [{\citenamefont {Miyake}\ \emph {et~al.}(2013)\citenamefont {Miyake},
  \citenamefont {Siviloglou}, \citenamefont {Kennedy}, \citenamefont {Burton},\
  and\ \citenamefont {Ketterle}}]{miyake_realizing_2013}%
  \BibitemOpen
  \bibfield  {author} {\bibinfo {author} {\bibfnamefont {H.}~\bibnamefont
  {Miyake}}, \bibinfo {author} {\bibfnamefont {G.~A.}\ \bibnamefont
  {Siviloglou}}, \bibinfo {author} {\bibfnamefont {C.~J.}\ \bibnamefont
  {Kennedy}}, \bibinfo {author} {\bibfnamefont {W.~C.}\ \bibnamefont
  {Burton}},\ and\ \bibinfo {author} {\bibfnamefont {W.}~\bibnamefont
  {Ketterle}},\ }\bibfield  {title} {\bibinfo {title} {Realizing the {{Harper
  Hamiltonian}} with {{Laser-Assisted Tunneling}} in {{Optical Lattices}}},\
  }\href@noop {} {\bibfield  {journal} {\bibinfo  {journal} {Phys. Rev. Lett.}\
  }\textbf {\bibinfo {volume} {111}},\ \bibinfo {pages} {185302} (\bibinfo
  {year} {2013})}\BibitemShut {NoStop}%
\bibitem [{\citenamefont {Aidelsburger}\ \emph {et~al.}(2013)\citenamefont
  {Aidelsburger}, \citenamefont {Atala}, \citenamefont {Lohse}, \citenamefont
  {Barreiro}, \citenamefont {Paredes},\ and\ \citenamefont
  {Bloch}}]{aidelsburger_realization_2013}%
  \BibitemOpen
  \bibfield  {author} {\bibinfo {author} {\bibfnamefont {M.}~\bibnamefont
  {Aidelsburger}}, \bibinfo {author} {\bibfnamefont {M.}~\bibnamefont {Atala}},
  \bibinfo {author} {\bibfnamefont {M.}~\bibnamefont {Lohse}}, \bibinfo
  {author} {\bibfnamefont {J.~T.}\ \bibnamefont {Barreiro}}, \bibinfo {author}
  {\bibfnamefont {B.}~\bibnamefont {Paredes}},\ and\ \bibinfo {author}
  {\bibfnamefont {I.}~\bibnamefont {Bloch}},\ }\bibfield  {title} {\bibinfo
  {title} {Realization of the {{Hofstadter Hamiltonian}} with {{Ultracold
  Atoms}} in {{Optical Lattices}}},\ }\href@noop {} {\bibfield  {journal}
  {\bibinfo  {journal} {Phys. Rev. Lett.}\ }\textbf {\bibinfo {volume} {111}},\
  \bibinfo {pages} {185301} (\bibinfo {year} {2013})}\BibitemShut {NoStop}%
\bibitem [{\citenamefont {Jotzu}\ \emph {et~al.}(2014)\citenamefont {Jotzu},
  \citenamefont {Messer}, \citenamefont {Desbuquois}, \citenamefont {Lebrat},
  \citenamefont {Uehlinger}, \citenamefont {Greif},\ and\ \citenamefont
  {Esslinger}}]{jotzu_experimental_2014}%
  \BibitemOpen
  \bibfield  {author} {\bibinfo {author} {\bibfnamefont {G.}~\bibnamefont
  {Jotzu}}, \bibinfo {author} {\bibfnamefont {M.}~\bibnamefont {Messer}},
  \bibinfo {author} {\bibfnamefont {R.}~\bibnamefont {Desbuquois}}, \bibinfo
  {author} {\bibfnamefont {M.}~\bibnamefont {Lebrat}}, \bibinfo {author}
  {\bibfnamefont {T.}~\bibnamefont {Uehlinger}}, \bibinfo {author}
  {\bibfnamefont {D.}~\bibnamefont {Greif}},\ and\ \bibinfo {author}
  {\bibfnamefont {T.}~\bibnamefont {Esslinger}},\ }\bibfield  {title} {\bibinfo
  {title} {Experimental realization of the topological {{Haldane}} model with
  ultracold fermions},\ }\href@noop {} {\bibfield  {journal} {\bibinfo
  {journal} {Nature}\ }\textbf {\bibinfo {volume} {515}},\ \bibinfo {pages}
  {237} (\bibinfo {year} {2014})}\BibitemShut {NoStop}%
\bibitem [{\citenamefont {{Cohen-Tannoudji}}\ \emph {et~al.}(1998)\citenamefont
  {{Cohen-Tannoudji}}, \citenamefont {{Dupont-Roc}},\ and\ \citenamefont
  {Grynberg}}]{cohen-tannoudji_atom-photon_1998}%
  \BibitemOpen
  \bibfield  {author} {\bibinfo {author} {\bibfnamefont {C.}~\bibnamefont
  {{Cohen-Tannoudji}}}, \bibinfo {author} {\bibfnamefont {J.}~\bibnamefont
  {{Dupont-Roc}}},\ and\ \bibinfo {author} {\bibfnamefont {G.}~\bibnamefont
  {Grynberg}},\ }\href@noop {} {\emph {\bibinfo {title} {Atom-Photon
  Interactions: Basic Processes and Applications}}}\ (\bibinfo  {publisher}
  {John Wiley \& Sons},\ \bibinfo {year} {1998})\BibitemShut {NoStop}%
\bibitem [{Note1()}]{Note1}%
  \BibitemOpen
  \bibinfo {note} {An earlier version of this manuscript (arXiv:2412.15038v2)
  used a slightly more involved configuration with $\alpha _+=1+\protect \frac
  {1}{2}\left [ \cos (kx)+\cos (ky)\right ]$ and $\alpha _-=\sin \left
  [\protect \frac {k}{2}(x-y)\right ]-{\protect \rm i}\sin \left [\protect
  \frac {k}{2}(x+y)\right ]$. This configuration was subsequently explored in
  detail in \cite {burba_two_2025}.}\BibitemShut {Stop}%
\bibitem [{\citenamefont {Dalibard}\ \emph {et~al.}(2011)\citenamefont
  {Dalibard}, \citenamefont {Gerbier}, \citenamefont {Juzeli{\=u}nas},\ and\
  \citenamefont {{\"O}hberg}}]{dalibard_colloquium_2011}%
  \BibitemOpen
  \bibfield  {author} {\bibinfo {author} {\bibfnamefont {J.}~\bibnamefont
  {Dalibard}}, \bibinfo {author} {\bibfnamefont {F.}~\bibnamefont {Gerbier}},
  \bibinfo {author} {\bibfnamefont {G.}~\bibnamefont {Juzeli{\=u}nas}},\ and\
  \bibinfo {author} {\bibfnamefont {P.}~\bibnamefont {{\"O}hberg}},\ }\bibfield
   {title} {\bibinfo {title} {{\emph{Colloquium}} : {{Artificial}} gauge
  potentials for neutral atoms},\ }\href@noop {} {\bibfield  {journal}
  {\bibinfo  {journal} {Rev. Mod. Phys.}\ }\textbf {\bibinfo {volume} {83}},\
  \bibinfo {pages} {1523} (\bibinfo {year} {2011})}\BibitemShut {NoStop}%
\bibitem [{Sup()}]{SuppMat_FL_DS}%
  \BibitemOpen
  \href@noop {} {\bibinfo {title} {See {{Supplemental Material}} for a
  discussion on the degeneracy condition of optical flux lattices, the ideal
  {{Chern}} band character, the extension to a triangular dark-state lattice,
  and a study of the topological robustness of the ground band. {{It}} includes
  {{Refs}}. [41-42]}}\BibitemShut {NoStop}%
\bibitem [{\citenamefont {Juzeli{\=u}nas}\ \emph {et~al.}(2005)\citenamefont
  {Juzeli{\=u}nas}, \citenamefont {{\"O}hberg}, \citenamefont {Ruseckas},\ and\
  \citenamefont {Klein}}]{juzeliunas_effective_2005}%
  \BibitemOpen
  \bibfield  {author} {\bibinfo {author} {\bibfnamefont {G.}~\bibnamefont
  {Juzeli{\=u}nas}}, \bibinfo {author} {\bibfnamefont {P.}~\bibnamefont
  {{\"O}hberg}}, \bibinfo {author} {\bibfnamefont {J.}~\bibnamefont
  {Ruseckas}},\ and\ \bibinfo {author} {\bibfnamefont {A.}~\bibnamefont
  {Klein}},\ }\bibfield  {title} {\bibinfo {title} {Effective magnetic fields
  in degenerate atomic gases induced by light beams with orbital angular
  momenta},\ }\href@noop {} {\bibfield  {journal} {\bibinfo  {journal} {Phys.
  Rev. A}\ }\textbf {\bibinfo {volume} {71}},\ \bibinfo {pages} {053614}
  (\bibinfo {year} {2005})}\BibitemShut {NoStop}%
\bibitem [{\citenamefont {Juzeli{\=u}nas}\ \emph {et~al.}(2006)\citenamefont
  {Juzeli{\=u}nas}, \citenamefont {Ruseckas}, \citenamefont {{\"O}hberg},\ and\
  \citenamefont {Fleischhauer}}]{juzeliunas_light-induced_2006}%
  \BibitemOpen
  \bibfield  {author} {\bibinfo {author} {\bibfnamefont {G.}~\bibnamefont
  {Juzeli{\=u}nas}}, \bibinfo {author} {\bibfnamefont {J.}~\bibnamefont
  {Ruseckas}}, \bibinfo {author} {\bibfnamefont {P.}~\bibnamefont
  {{\"O}hberg}},\ and\ \bibinfo {author} {\bibfnamefont {M.}~\bibnamefont
  {Fleischhauer}},\ }\bibfield  {title} {\bibinfo {title} {Light-induced
  effective magnetic fields for ultracold atoms in planar geometries},\
  }\href@noop {} {\bibfield  {journal} {\bibinfo  {journal} {Phys. Rev. A}\
  }\textbf {\bibinfo {volume} {73}},\ \bibinfo {pages} {025602} (\bibinfo
  {year} {2006})}\BibitemShut {NoStop}%
\bibitem [{\citenamefont {Ruseckas}\ \emph {et~al.}(2005)\citenamefont
  {Ruseckas}, \citenamefont {Juzeli{\=u}nas}, \citenamefont {{\"O}hberg},\ and\
  \citenamefont {Fleischhauer}}]{ruseckas_non-abelian_2005}%
  \BibitemOpen
  \bibfield  {author} {\bibinfo {author} {\bibfnamefont {J.}~\bibnamefont
  {Ruseckas}}, \bibinfo {author} {\bibfnamefont {G.}~\bibnamefont
  {Juzeli{\=u}nas}}, \bibinfo {author} {\bibfnamefont {P.}~\bibnamefont
  {{\"O}hberg}},\ and\ \bibinfo {author} {\bibfnamefont {M.}~\bibnamefont
  {Fleischhauer}},\ }\bibfield  {title} {\bibinfo {title} {Non-{{Abelian Gauge
  Potentials}} for {{Ultracold Atoms}} with {{Degenerate Dark States}}},\
  }\href@noop {} {\bibfield  {journal} {\bibinfo  {journal} {Phys. Rev. Lett.}\
  }\textbf {\bibinfo {volume} {95}},\ \bibinfo {pages} {010404} (\bibinfo
  {year} {2005})}\BibitemShut {NoStop}%
\bibitem [{\citenamefont {Sommer}\ and\ \citenamefont
  {Cooper}(2025)}]{sommer_ideal_2025}%
  \BibitemOpen
  \bibfield  {author} {\bibinfo {author} {\bibfnamefont {O.~E.}\ \bibnamefont
  {Sommer}}\ and\ \bibinfo {author} {\bibfnamefont {N.~R.}\ \bibnamefont
  {Cooper}},\ }\bibfield  {title} {\bibinfo {title} {Ideal {{Optical Flux
  Lattices}}},\ }\href@noop {} {\bibfield  {journal} {\bibinfo  {journal}
  {arXiv:2509.01481}\ } (\bibinfo {year} {2025})}\BibitemShut {NoStop}%
\bibitem [{\citenamefont {Haldane}(1985)}]{haldane_many-particle_1985}%
  \BibitemOpen
  \bibfield  {author} {\bibinfo {author} {\bibfnamefont {F.~D.~M.}\
  \bibnamefont {Haldane}},\ }\bibfield  {title} {\bibinfo {title}
  {Many-{{Particle Translational Symmetries}} of {{Two-Dimensional Electrons}}
  at {{Rational Landau-Level Filling}}},\ }\href@noop {} {\bibfield  {journal}
  {\bibinfo  {journal} {Phys. Rev. Lett.}\ }\textbf {\bibinfo {volume} {55}},\
  \bibinfo {pages} {2095} (\bibinfo {year} {1985})}\BibitemShut {NoStop}%
\bibitem [{\citenamefont {Haldane}\ and\ \citenamefont
  {Rezayi}(1985)}]{haldane_periodic_1985}%
  \BibitemOpen
  \bibfield  {author} {\bibinfo {author} {\bibfnamefont {F.~D.~M.}\
  \bibnamefont {Haldane}}\ and\ \bibinfo {author} {\bibfnamefont {E.~H.}\
  \bibnamefont {Rezayi}},\ }\bibfield  {title} {\bibinfo {title} {Periodic
  {{Laughlin-Jastrow}} wave functions for the fractional quantized {{Hall}}
  effect},\ }\href@noop {} {\bibfield  {journal} {\bibinfo  {journal} {Phys.
  Rev. B}\ }\textbf {\bibinfo {volume} {31}},\ \bibinfo {pages} {2529}
  (\bibinfo {year} {1985})}\BibitemShut {NoStop}%
\bibitem [{\citenamefont {Read}\ and\ \citenamefont
  {Rezayi}(1996)}]{read_quasiholes_1996}%
  \BibitemOpen
  \bibfield  {author} {\bibinfo {author} {\bibfnamefont {N.}~\bibnamefont
  {Read}}\ and\ \bibinfo {author} {\bibfnamefont {E.}~\bibnamefont {Rezayi}},\
  }\bibfield  {title} {\bibinfo {title} {Quasiholes and fermionic zero modes of
  paired fractional quantum {{Hall}} states: {{The}} mechanism for
  non-{{Abelian}} statistics},\ }\href@noop {} {\bibfield  {journal} {\bibinfo
  {journal} {Phys. Rev. B}\ }\textbf {\bibinfo {volume} {54}},\ \bibinfo
  {pages} {16864} (\bibinfo {year} {1996})}\BibitemShut {NoStop}%
\bibitem [{\citenamefont {Ledwith}\ \emph {et~al.}(2023)\citenamefont
  {Ledwith}, \citenamefont {Vishwanath},\ and\ \citenamefont
  {Parker}}]{ledwith_vortexability_2023}%
  \BibitemOpen
  \bibfield  {author} {\bibinfo {author} {\bibfnamefont {P.~J.}\ \bibnamefont
  {Ledwith}}, \bibinfo {author} {\bibfnamefont {A.}~\bibnamefont
  {Vishwanath}},\ and\ \bibinfo {author} {\bibfnamefont {D.~E.}\ \bibnamefont
  {Parker}},\ }\bibfield  {title} {\bibinfo {title} {Vortexability: {{A}}
  unifying criterion for ideal fractional {{Chern}} insulators},\ }\href@noop
  {} {\bibfield  {journal} {\bibinfo  {journal} {Phys. Rev. B}\ }\textbf
  {\bibinfo {volume} {108}},\ \bibinfo {pages} {205144} (\bibinfo {year}
  {2023})}\BibitemShut {NoStop}%
\bibitem [{\citenamefont {Halperin}(1982)}]{halperin_quantized_1982}%
  \BibitemOpen
  \bibfield  {author} {\bibinfo {author} {\bibfnamefont {B.~I.}\ \bibnamefont
  {Halperin}},\ }\bibfield  {title} {\bibinfo {title} {Quantized {{Hall}}
  conductance, current-carrying edge states, and the existence of extended
  states in a two-dimensional disordered potential},\ }\href@noop {} {\bibfield
   {journal} {\bibinfo  {journal} {Phys. Rev. B}\ }\textbf {\bibinfo {volume}
  {25}},\ \bibinfo {pages} {2185} (\bibinfo {year} {1982})}\BibitemShut
  {NoStop}%
\bibitem [{\citenamefont {Hatsugai}(1993)}]{hatsugai_chern_1993}%
  \BibitemOpen
  \bibfield  {author} {\bibinfo {author} {\bibfnamefont {Y.}~\bibnamefont
  {Hatsugai}},\ }\bibfield  {title} {\bibinfo {title} {Chern number and edge
  states in the integer quantum {{Hall}} effect},\ }\href@noop {} {\bibfield
  {journal} {\bibinfo  {journal} {Phys. Rev. Lett.}\ }\textbf {\bibinfo
  {volume} {71}},\ \bibinfo {pages} {3697} (\bibinfo {year}
  {1993})}\BibitemShut {NoStop}%
\bibitem [{\citenamefont {Madison}\ \emph {et~al.}(2000)\citenamefont
  {Madison}, \citenamefont {Chevy}, \citenamefont {Wohlleben},\ and\
  \citenamefont {Dalibard}}]{madison_vortex_2000}%
  \BibitemOpen
  \bibfield  {author} {\bibinfo {author} {\bibfnamefont {K.~W.}\ \bibnamefont
  {Madison}}, \bibinfo {author} {\bibfnamefont {F.}~\bibnamefont {Chevy}},
  \bibinfo {author} {\bibfnamefont {W.}~\bibnamefont {Wohlleben}},\ and\
  \bibinfo {author} {\bibfnamefont {J.}~\bibnamefont {Dalibard}},\ }\bibfield
  {title} {\bibinfo {title} {Vortex formation in a stirred {{Bose-Einstein}}
  condensate},\ }\href@noop {} {\bibfield  {journal} {\bibinfo  {journal}
  {Phys. Rev. Lett.}\ }\textbf {\bibinfo {volume} {84}},\ \bibinfo {pages}
  {806} (\bibinfo {year} {2000})}\BibitemShut {NoStop}%
\bibitem [{\citenamefont {{Abo-Shaeer}}\ \emph {et~al.}(2001)\citenamefont
  {{Abo-Shaeer}}, \citenamefont {Raman}, \citenamefont {Vogels},\ and\
  \citenamefont {Ketterle}}]{abo-shaeer_observation_2001}%
  \BibitemOpen
  \bibfield  {author} {\bibinfo {author} {\bibfnamefont {J.~R.}\ \bibnamefont
  {{Abo-Shaeer}}}, \bibinfo {author} {\bibfnamefont {C.}~\bibnamefont {Raman}},
  \bibinfo {author} {\bibfnamefont {J.~M.}\ \bibnamefont {Vogels}},\ and\
  \bibinfo {author} {\bibfnamefont {W.}~\bibnamefont {Ketterle}},\ }\bibfield
  {title} {\bibinfo {title} {Observation of vortex lattices in
  {{Bose-Einstein}} condensates},\ }\href@noop {} {\bibfield  {journal}
  {\bibinfo  {journal} {Science}\ }\textbf {\bibinfo {volume} {292}},\ \bibinfo
  {pages} {476} (\bibinfo {year} {2001})}\BibitemShut {NoStop}%
\bibitem [{\citenamefont {Schweikhard}\ \emph {et~al.}(2004)\citenamefont
  {Schweikhard}, \citenamefont {Coddington}, \citenamefont {Engels},
  \citenamefont {Mogendorff},\ and\ \citenamefont
  {Cornell}}]{schweikhard_rapidly_2004}%
  \BibitemOpen
  \bibfield  {author} {\bibinfo {author} {\bibfnamefont {V.}~\bibnamefont
  {Schweikhard}}, \bibinfo {author} {\bibfnamefont {I.}~\bibnamefont
  {Coddington}}, \bibinfo {author} {\bibfnamefont {P.}~\bibnamefont {Engels}},
  \bibinfo {author} {\bibfnamefont {V.~P.}\ \bibnamefont {Mogendorff}},\ and\
  \bibinfo {author} {\bibfnamefont {E.~A.}\ \bibnamefont {Cornell}},\
  }\bibfield  {title} {\bibinfo {title} {Rapidly {{Rotating Bose-Einstein
  Condensates}} in and near the {{Lowest Landau Level}}},\ }\href@noop {}
  {\bibfield  {journal} {\bibinfo  {journal} {Phys. Rev. Lett.}\ }\textbf
  {\bibinfo {volume} {92}} (\bibinfo {year} {2004})}\BibitemShut {NoStop}%
\bibitem [{\citenamefont {Mukherjee}\ \emph {et~al.}(2022)\citenamefont
  {Mukherjee}, \citenamefont {Shaffer}, \citenamefont {Patel}, \citenamefont
  {Yan}, \citenamefont {Wilson}, \citenamefont {Cr{\'e}pel}, \citenamefont
  {Fletcher},\ and\ \citenamefont
  {Zwierlein}}]{mukherjee_crystallization_2022}%
  \BibitemOpen
  \bibfield  {author} {\bibinfo {author} {\bibfnamefont {B.}~\bibnamefont
  {Mukherjee}}, \bibinfo {author} {\bibfnamefont {A.}~\bibnamefont {Shaffer}},
  \bibinfo {author} {\bibfnamefont {P.~B.}\ \bibnamefont {Patel}}, \bibinfo
  {author} {\bibfnamefont {Z.}~\bibnamefont {Yan}}, \bibinfo {author}
  {\bibfnamefont {C.~C.}\ \bibnamefont {Wilson}}, \bibinfo {author}
  {\bibfnamefont {V.}~\bibnamefont {Cr{\'e}pel}}, \bibinfo {author}
  {\bibfnamefont {R.~J.}\ \bibnamefont {Fletcher}},\ and\ \bibinfo {author}
  {\bibfnamefont {M.}~\bibnamefont {Zwierlein}},\ }\bibfield  {title} {\bibinfo
  {title} {Crystallization of bosonic quantum {{Hall}} states in a rotating
  quantum gas},\ }\href@noop {} {\bibfield  {journal} {\bibinfo  {journal}
  {Nature}\ }\textbf {\bibinfo {volume} {601}},\ \bibinfo {pages} {58}
  (\bibinfo {year} {2022})}\BibitemShut {NoStop}%
\bibitem [{\citenamefont {Baur}\ and\ \citenamefont
  {Cooper}(2013)}]{baur_adiabatic_2013}%
  \BibitemOpen
  \bibfield  {author} {\bibinfo {author} {\bibfnamefont {S.~K.}\ \bibnamefont
  {Baur}}\ and\ \bibinfo {author} {\bibfnamefont {N.~R.}\ \bibnamefont
  {Cooper}},\ }\bibfield  {title} {\bibinfo {title} {Adiabatic preparation of
  vortex lattices},\ }\href@noop {} {\bibfield  {journal} {\bibinfo  {journal}
  {Phys. Rev. A}\ }\textbf {\bibinfo {volume} {88}},\ \bibinfo {pages} {033603}
  (\bibinfo {year} {2013})}\BibitemShut {NoStop}%
\bibitem [{\citenamefont {Read}\ and\ \citenamefont
  {Cooper}(2003)}]{read_free_2003}%
  \BibitemOpen
  \bibfield  {author} {\bibinfo {author} {\bibfnamefont {N.}~\bibnamefont
  {Read}}\ and\ \bibinfo {author} {\bibfnamefont {N.~R.}\ \bibnamefont
  {Cooper}},\ }\bibfield  {title} {\bibinfo {title} {Free expansion of
  lowest-{{Landau-level}} states of trapped atoms: {{A}} wave-function
  microscope},\ }\href@noop {} {\bibfield  {journal} {\bibinfo  {journal}
  {Phys. Rev. A}\ }\textbf {\bibinfo {volume} {68}},\ \bibinfo {pages} {035601}
  (\bibinfo {year} {2003})}\BibitemShut {NoStop}%
\bibitem [{\citenamefont {Wang}\ \emph {et~al.}(2021)\citenamefont {Wang},
  \citenamefont {Cano}, \citenamefont {Millis}, \citenamefont {Liu},\ and\
  \citenamefont {Yang}}]{wang_exact_2021}%
  \BibitemOpen
  \bibfield  {author} {\bibinfo {author} {\bibfnamefont {J.}~\bibnamefont
  {Wang}}, \bibinfo {author} {\bibfnamefont {J.}~\bibnamefont {Cano}}, \bibinfo
  {author} {\bibfnamefont {A.~J.}\ \bibnamefont {Millis}}, \bibinfo {author}
  {\bibfnamefont {Z.}~\bibnamefont {Liu}},\ and\ \bibinfo {author}
  {\bibfnamefont {B.}~\bibnamefont {Yang}},\ }\bibfield  {title} {\bibinfo
  {title} {Exact {{Landau Level Description}} of {{Geometry}} and
  {{Interaction}} in a {{Flatband}}},\ }\href@noop {} {\bibfield  {journal}
  {\bibinfo  {journal} {Phys. Rev. Lett.}\ }\textbf {\bibinfo {volume} {127}},\
  \bibinfo {pages} {246403} (\bibinfo {year} {2021})}\BibitemShut {NoStop}%
\bibitem [{\citenamefont {Satoor}\ \emph {et~al.}(2021)\citenamefont {Satoor},
  \citenamefont {Fabre}, \citenamefont {Bouhiron}, \citenamefont {Evrard},
  \citenamefont {Lopes},\ and\ \citenamefont
  {Nascimbene}}]{satoor_partitioning_2021}%
  \BibitemOpen
  \bibfield  {author} {\bibinfo {author} {\bibfnamefont {T.}~\bibnamefont
  {Satoor}}, \bibinfo {author} {\bibfnamefont {A.}~\bibnamefont {Fabre}},
  \bibinfo {author} {\bibfnamefont {J.-B.}\ \bibnamefont {Bouhiron}}, \bibinfo
  {author} {\bibfnamefont {A.}~\bibnamefont {Evrard}}, \bibinfo {author}
  {\bibfnamefont {R.}~\bibnamefont {Lopes}},\ and\ \bibinfo {author}
  {\bibfnamefont {S.}~\bibnamefont {Nascimbene}},\ }\bibfield  {title}
  {\bibinfo {title} {Partitioning dysprosium's electronic spin to reveal
  entanglement in nonclassical states},\ }\href@noop {} {\bibfield  {journal}
  {\bibinfo  {journal} {Phys. Rev. Research}\ }\textbf {\bibinfo {volume}
  {3}},\ \bibinfo {pages} {3,} (\bibinfo {year} {2021})}\BibitemShut {NoStop}%
\bibitem [{\citenamefont {{A. Kramida}}\ \emph {et~al.}(2025)\citenamefont {{A.
  Kramida}}, \citenamefont {{Yu. Ralchenko}}, \citenamefont {{J. Reader}},\
  and\ \citenamefont {{NIST ASD Team}}}]{NIST_ASD}%
  \BibitemOpen
  \bibfield  {author} {\bibinfo {author} {\bibnamefont {{A. Kramida}}},
  \bibinfo {author} {\bibnamefont {{Yu. Ralchenko}}}, \bibinfo {author}
  {\bibnamefont {{J. Reader}}},\ and\ \bibinfo {author} {\bibnamefont {{NIST
  ASD Team}}},\ }\href@noop {} {\bibfield  {journal} {\bibinfo  {journal} {NIST
  Atomic Spectra Database, Institute of Standards and Technology, Gaithersburg,
  MD}\ } (\bibinfo {year} {2025})}\BibitemShut {NoStop}%
\bibitem [{\citenamefont {Li}\ \emph {et~al.}(2017)\citenamefont {Li},
  \citenamefont {Wyart}, \citenamefont {Dulieu}, \citenamefont {Nascimbene},\
  and\ \citenamefont {Lepers}}]{li_optical_2017}%
  \BibitemOpen
  \bibfield  {author} {\bibinfo {author} {\bibfnamefont {H.}~\bibnamefont
  {Li}}, \bibinfo {author} {\bibfnamefont {J.-F.}\ \bibnamefont {Wyart}},
  \bibinfo {author} {\bibfnamefont {O.}~\bibnamefont {Dulieu}}, \bibinfo
  {author} {\bibfnamefont {S.}~\bibnamefont {Nascimbene}},\ and\ \bibinfo
  {author} {\bibfnamefont {M.}~\bibnamefont {Lepers}},\ }\bibfield  {title}
  {\bibinfo {title} {Optical trapping of ultracold dysprosium atoms: transition
  probabilities, dynamic dipole polarizabilities and van der waals $C_6$
  coefficients},\ }\href@noop {} {\bibfield  {journal} {\bibinfo  {journal} {J.
  Phys. B At. Mol. Opt. Phys.}\ }\textbf {\bibinfo {volume} {50}},\ \bibinfo
  {pages} {014005} (\bibinfo {year} {2017})}\BibitemShut {NoStop}%
\bibitem [{\citenamefont {Roy}(2014)}]{roy_band_2014}%
  \BibitemOpen
  \bibfield  {author} {\bibinfo {author} {\bibfnamefont {R.}~\bibnamefont
  {Roy}},\ }\bibfield  {title} {\bibinfo {title} {Band geometry of fractional
  topological insulators},\ }\href@noop {} {\bibfield  {journal} {\bibinfo
  {journal} {Phys. Rev. B}\ }\textbf {\bibinfo {volume} {90}},\ \bibinfo
  {pages} {165139} (\bibinfo {year} {2014})}\BibitemShut {NoStop}%
\bibitem [{\citenamefont {Estienne}\ \emph {et~al.}(2023)\citenamefont
  {Estienne}, \citenamefont {Regnault},\ and\ \citenamefont
  {Cr{\'e}pel}}]{estienne_ideal_2023}%
  \BibitemOpen
  \bibfield  {author} {\bibinfo {author} {\bibfnamefont {B.}~\bibnamefont
  {Estienne}}, \bibinfo {author} {\bibfnamefont {N.}~\bibnamefont {Regnault}},\
  and\ \bibinfo {author} {\bibfnamefont {V.}~\bibnamefont {Cr{\'e}pel}},\
  }\bibfield  {title} {\bibinfo {title} {Ideal {{Chern}} bands as {{Landau}}
  levels in curved space},\ }\href@noop {} {\bibfield  {journal} {\bibinfo
  {journal} {Phys. Rev. Res.}\ }\textbf {\bibinfo {volume} {5}},\ \bibinfo
  {pages} {L032048} (\bibinfo {year} {2023})}\BibitemShut {NoStop}%
\end{thebibliography}

%


\newpage
\vskip 1cm
\centerline{\textbf{END MATTER}}
\vskip 5mm

\centerline{\textbf{1. Dark state of a $\Lambda$ transition}}
\vskip 2mm

In this paragraph, we consider the $\Lambda$ transition sketched in \fig{fig:bands}a, adding the possibility of a non-zero Raman detuning, with $\Delta_+\equiv \omega_1-\omega_+$ differing slightly from  $\Delta_-\equiv \omega_2-\omega_-$. In the dressed atom picture \cite{cohen-tannoudji_atom-photon_1998}, denoting $n_\pm$ the number of photons in each light wave, the laser excitation couples the three states
\begin{equation}
|g_+,n_++1,n_-\rangle \leftrightarrow |e,n_+,n_-\rangle \leftrightarrow |g_-,n_+,n_-+1\rangle
\end{equation}
with the space-dependent Rabi frequencies $\kappa_\pm(\bs r)$. The laser coupling reads in this 3-dimensional space
\begin{equation}
\hat W(\bs r)=\hbar \begin{pmatrix}
+\delta/2 & \kappa_+^* & 0 \\ \kappa_+ & -\Delta-{\rm i}\Gamma/2 & \kappa_- \\ 0 & \kappa_-^* & -\delta/2
\end{pmatrix}
\label{eq:matrix_W}
\end{equation}
where we have introduced the average detuning $\Delta=(\Delta_++\Delta_-)/2$ and the detuning from the Raman resonance $\delta=\Delta_+-\Delta_-$. We also added an imaginary component $-{\rm i}\Gamma/2$ to the energy of the excited state $e$ to take into account its finite radiative lifetime $\Gamma^{-1}$. For $\delta=0$, the state $|{\cal D}(\bs r)\rangle \propto \kappa_-(\bs r) |g_+\rangle -\kappa_+(\bs r)|g_-\rangle$ is an eigenstate with energy 0 of the matrix $\hat W$.

In the limit $|\Delta|\gg |\kappa_\pm|,|\delta|,\Gamma$, we can eliminate the excited state adiabatically and obtain an effective Hamiltonian acting in the $\{|g_+\rangle,|g_-\rangle\}$ basis:
\begin{equation}
\hat V(\bs r)=\frac{1}{2}\hbar \delta \hat \sigma_z +\frac{\hbar}{\Delta}\begin{pmatrix}|\kappa_+|^2 & \kappa_+^* \kappa_- \\ \kappa_+ \kappa_-^* &  |\kappa_-|^2 \end{pmatrix},
\end{equation}
where $\hat\sigma_z$ is the Pauli matrix. We recover the light coupling formula (\ref{eq:V_DS}) upon the identification $\kappa_\pm=\sqrt{\Delta/V_0\hbar}\,\alpha_\pm$. In the main text, we restrict to the case where the Raman condition is fulfilled ($\delta=0$), which directly leads to the structure (\ref{eq:V_DS}) for the atom-laser coupling. In Appendix 3  we evaluate the effect of the finite lifetime of the excited state by solving the eigenvalue problem for the full Hamiltonian $\hat  {\bs p}^2/2m+\hat W(\bs r)$. The effect of a non-zero Raman detuning $\delta$ is addressed in \cite{SuppMat_FL_DS}.

\vskip 5mm


\centerline{\textbf{2. Admixture with the bright state}}
\vskip 2mm

In this paragraph, we discuss the admixture of the quantum states of the ground band of the dark optical lattice with the bright subspace of the $\Lambda$ transition.

\begin{figure}[t]
  \begin{center}
     \includegraphics[trim=12 16 0 0,scale=0.92]{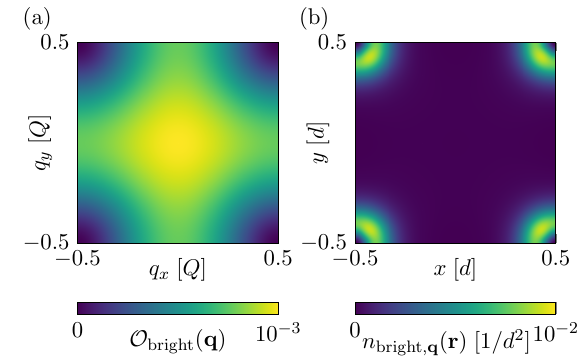}
  \end{center}
  \caption{(a) Overlap $\mathcal{O}_{\text{bright}}(\qbold)$ of a given Bloch state of quasimomentum $\qbold$ of the lowest band with the bright subspace, computed for $V_0=100\,\hbar\omc$. (b) Density probability in the bright subspace $n_{\text{bright},\qbold}(\rbold)$ for the quasi-momentum $\qbold=0$.}
  \label{fig:admixture}
\end{figure}

In a first step, we consider the overlap $\mathcal{O}_{\text{bright}}(\qbold)$ of a given Bloch state of quasimomentum $\qbold$ with the bright subspace, plotted in \fig{fig:admixture}a. The bright state admixture remains very small for all values of quasi-momentum. We checked that it decays with the lattice depth as  $\mathcal{O}_{\text{bright}}(\qbold)\propto \hbar\omega_c/V_0$, in agreement with the  first correction to the Born-Oppenheimer approximation. 

Remarkably, we find that the bright state admixture cancels at the corner $\qbold=(\pi/d,\pi/d)$ of the first Brillouin zone. This behavior can be explained by the fact that the position-dependent dark state
\begin{align*}
&|{\cal D}(\bs r)\rangle \propto  \; \alpha_-(\bs r) |g_+\rangle -\alpha_+(\bs r)|g_-\rangle\\
&\propto   \left[\E^{\I \frac{k}{2}(x-y)}+\E^{\I \frac{k}{2}(-x+y)}+\E^{\I \frac{k}{2}(x+y)}+\E^{\I \frac{k}{2}(-x-y)}\right]|g_+\rangle\\
& +\left[\E^{\I \frac{k}{2}(x-y)}-\E^{\I \frac{k}{2}(-x+y)}+\I \,\E^{\I \frac{k}{2}(x+y)}-\I \,\E^{\I \frac{k}{2}(-x-y)}\right]|g_-\rangle
\end{align*}
is an exact eigenstate of the Hamiltonian. Indeed, it is by construction the zero-energy eigenstate of the light shift operator $\hat V(\rbold)$ for all positions $\rbold$. Since it only involves plane waves of momentum $(k/2)(\pm\ebold_x\pm\ebold_y)$, it is also an exact eigenstate of the kinetic energy operator with  energy $E_r$, where $E_r\equiv\hbar^2k^2/(4m)$ is the recoil energy for this transition (recall that $k=k_{1}/\sqrt2$). Finally, it corresponds to  a quasi-momentum $\qbold=(\pi/d,\pi/d)$, hence it matches with the quasi-momentum state of the lowest Bloch band at the corner of the first Brillouin zone.


In a second step, we study the locations in real space where the admixture of a Bloch state of quasimomentum $\qbold\neq (\pi/d,\pi/d)$ is maximal.  For this, we compute numerically the position-resolved density probability for a given quasi-momentum state $\qbold$ of the lowest band to be found in the bright subspace :
\[
n_{\text{bright},\qbold}(\rbold)=|\langle\psi_\qbold(\rbold)|{\cal B}(\bs r)\rangle|^2.
\]
We find that this admixture predominantly occurs  in the vicinity of the unit cell corner $d(\ebold_x+\ebold_y)/2$ (see \fig{fig:admixture}b). It can be explained by the fact that these corners are the locations where the Wannier functions of the lowest energy level for the bright state are maximal. Hence for an atom moving in the LLL associated to the dark state, the probability to have a non-adiabatic transfer to these Wannier functions is maximal at the unit cell corners. 



\newpage

\centerline{\textbf{3. Implementation with dysprosium atoms}}
\vskip 2mm

\begin{figure}[t]
  \begin{center}
  \includegraphics[scale=0.94,trim=14 22 0 3]{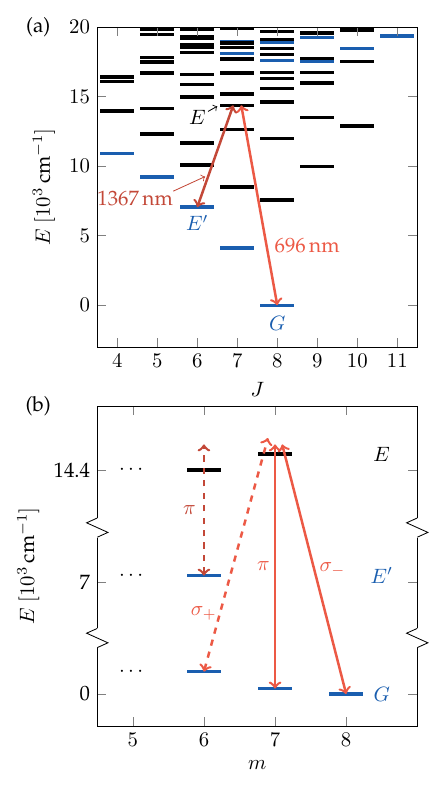}
  \end{center}
  \caption{(a) Level scheme of atomic dysprosium up to a 20\,000\,cm\textsuperscript{-1} energy \cite{NIST_ASD}. Even (odd) parity levels are shown in blue (black, respectively). We label the three energy levels $G$, $E$ and $E'$ considered in appendix 3, together with the two optical transitions connecting them. (b) Scheme of the magnetic levels considered in appendix 3. The three levels $\ket{G,m=8}$, $\ket{G,m=7}$ and $\ket{E,m=7}$ forming the $\Lambda$ configuration are coupled using off-resonant light at \SI{696}{\nano\meter} with polarization $\pi$ and $\sigma_-$ with respect to the applied quantization magnetic field. The dashed arrows represent additional light couplings used to lift the energy of the undesired states $\ket{G,m=6}$ and $\ket{E,m=6}$.}
  \label{fig:dy}
\end{figure}

The dark optical lattice studied in the main text involves a triple of atomic states in a $\Lambda$ configuration. Lanthanide atoms, which exhibit a rich electronic structure with isolated narrow optical transitions, are well suited for isolating such a triple of states. We take here the example of dysprosium atoms, whose electronic structure is shown in \fig{fig:dy}a. For sake of simplicity, we consider bosonic atoms, of nuclear spin $I=0$, but the scheme can be straightforwardly extended to fermionic isotopes. 

The ground electronic level $G$ exhibits an angular momentum $J_G=8$. For the two ground states of the $\Lambda$ configuration, we consider the two magnetic states $|g_-\rangle \equiv |G,m=J_G-1\rangle$, $|g_+\rangle \equiv |G,m=J_G\rangle$. The excited level is defined as the streched state $|e\rangle\equiv|E,m=J_E\rangle$ of the excited level of energy $\SI{14 368}{cm^{-1}}$ above the ground state, and angular momentum $J_E=7$. These three states can be coupled using coherent light of wavelength $\lambda=\SI{696}{\nano\meter}$ with $\pi$ and $\sigma_-$ polarization. As discussed in the main text, this light must be blue-detuned with respect to the optical resonance, withnece a detuning $\Delta$ much larger than the light shift $V_0$, the cyclotron frequency $\omega_{\text{c}}\simeq h\times\SI{1.6}{\kilo\hertz}$ and the excited-state inverse lifetime $\Gamma\simeq\SI{94}{\milli\second^{-1}}$.

The Landau level structure, with quasi-flat energy bands, can be deformed due to the presence of additional magnetic levels, such as the ground-state level $\ket{G,m=J_G-2}$ and the excited state $\ket{E,m=J_E-1}$. Those states much be shifted in energy using additional laser beams in order to suppress their influence on the low-lying bandstructure. 

More specifically, the state $\ket{G,m=J_G-2}$ must be shifted to an energy $\Delta_1\gg E_{\text{rec}}$, which can easily be done using a blue-detuned light beam at $\SI{696}{\nano\meter}$ with $\sigma_+$ polarization,  without affecting the states $\ket{g_\pm}$ \cite{satoor_partitioning_2021}.

The state $\ket{E,m=J_E-1}$ is off-resonantly coupled to the state $\ket{g_-}$ by the $\sigma_-$-polarized light at $\SI{696}{\nano\meter}$, leading to an additional light shift that can potentially deform the low-lying bands. This effect can be made negligible by shifting the energy of this state by a value $\Delta_2<0$, with $|\Delta_2|$ largely exceeding the detuning $\Delta$. This can be achieved by off-resonant coupling  this state to another electronic level $\ket{E',m=J_{E'}}$, of energy $\SI{7051}{\centi\meter^{-1}}$ above the ground state and angular momentum $J_{E'}=6$. By using a light polarization $\pi$, the excited level $\ket{e}$ remains unaffected. 

Finally, we estimate the photon scattering rate induced by the off-resonant population of excited levels. For this, we compute the bandstructure including the levels $\ket{g_-}$, $\ket{g_+}$, $\ket{e}$, $\ket{G,m=J_G-2}$ and $\ket{E,m=J_E-1}$, including an imaginary part $\I\Gamma/2$ in the energy of the excited states. We obtain a bandstructure with quasi-flat energy bands for $V_0=100\,\hbar\omega_{\text{c}}$, a detuning $\Delta=25\,V_0\simeq2\pi\times\SI{4}{\mega\hertz}$, and light shifts of undesired states $\Delta_1=2\pi\times\SI{100}{\kilo\hertz}$ and $\Delta_2=-100\,\Delta\simeq2\pi\times\SI{400}{\mega\hertz}$. For these parameters, we compute a mean scattering rate $\Gamma_{\text{sc}}\simeq\SI{1}{\second^{-1}}$ from the imaginary part of the ground-band eigenstates. We note that, given the small value of the detuning $\Delta$, the contribution to the light shift and scattering rate from the other transitions is negligible \cite{li_optical_2017}.   The low value of the scattering rate makes this protocol promising for the investigation of interacting many-body ground states in topological bands.

Equally promising results are obtained for the fermionic isotope 87 of Strontium, with the lattice light tuned to the intercombination line $^1$S$_0$ - $^3$P$_1$, more
specifically using the states $\ket{g_-}=\ket{G,F=9/2,m=7/2}$,  $\ket{g_+}=\ket{G,F=9/2,m=9/2}$ and  $\ket{e}=\ket{E,F'=7/2,m'=7/2}$.

\cleardoublepage

\setcounter{equation}{0}
\setcounter{figure}{0}
\setcounter{table}{0}
\setcounter{page}{1}
\makeatletter
\renewcommand{\theequation}{S\arabic{equation}}
\renewcommand{\thefigure}{S\arabic{figure}}


\centerline{\textbf{
Supplemental Material
 }}
 
 \medskip
 
 \centerline{\textbf{
Emergence of a Landau level structure
 }}
 \centerline{\textbf{
in dark optical lattices
 }}
 
 \bigskip
 \bigskip

\centerline{\textbf{S1. The non-degeneracy condition of OFLs}}
\vskip 2mm
In its simplest version \cite{cooper_optical_2011}, an optical flux lattice (OFL) consists in a spatially periodic coupling between two internal states $|a\rangle$ and $|b\rangle$ (not necessarily the two states $|g_+\rangle$ and $|g_-\rangle$ introduced in the main text). This coupling can be written $\hat V=\epsilon_0\hat 1+ \epsilon_1\, \bs n\cdot \bs {\hat\sigma}$, where the real functions $\epsilon_0(\bs r)$, $\epsilon_1(\bs r)$ and the unit vector $\bs n(\bs r)$ are spatially periodic, and the $\hat \sigma_j$ ($j=x,y,z$) are the Pauli matrices. 

At any point in space, we can diagonalize $\hat V$ and obtain the two eigenstates $|\pm\rangle_{\bs n}$ with the energies $\epsilon_0\pm \epsilon_1$. The non-degeneracy condition stated in \cite{cooper_optical_2011} imposes that $\epsilon_1$ does not cancel at any point in the unit cell. Suppose for example that $\epsilon_1$ is strictly positive everywhere, and consider the Berry connection $\bs A(\bs r)=\I \hbar\  _{\bs n}\langle -|\bs \nabla\left( |-\rangle_{\bs n}\right)$ and the Berry curvature $B_z(\bs r)=\left( \bs \nabla \times \bs A\right)_z$ that emerge for an adiabatic following of the local ground state $|-\rangle_{\bs n}$. By definition of an OFL, the flux of $B_z$ across the unit cell should be non-zero. Using Stokes theorem, this flux is given by:
\begin{equation}
\int\hskip-3mm\int B_z \;{\D}^2r=\oint_{\cal C}\bs A\cdot \D\bs r+\mbox{singular contributions,}
\label{eq:Stokes}
\end{equation} 
where the contour ${\cal C}$ represents the edge of the unit cell. Since $\bs A$ is periodic on the lattice, its contour integral along the cell edge is zero and one is left with only the contribution of singularities \cite{cooper_optical_2011,juzeliunas_flux_2012}. One must therefore engineer these singularities such that their sum gives a non-zero result. 

Let us choose a gauge, for example $|-\rangle_{\bs n}=(-{\rm e}^{{-\rm i}\phi}\sin(\theta/2),\cos(\theta/2))^T$ where the spherical angles $(\theta,\phi)$ define the orientation of $\bs n$. The Berry connection then reads $\bs A=\hbar\,\sin^2(\theta/2) \,\bs \nabla \phi$, showing that a singularity of $\bs A$ can appear where $\phi$ is ill-defined. This occurs at the zeroes of the complex periodic function $V_{ab}=\epsilon_1\,{\rm e}^{-{\rm i}\phi}\,\cos \theta$, hence at places where $\theta=0$ or $\pi$. More precisely, a point where $\theta=\pi$, i.e. $V_{aa}-\epsilon_0=\epsilon_1\cos\theta <0$ will lead to a singular contribution since $\bs A=\hbar \,\bs \nabla \phi$ at this point, whereas $\bs A$ will vanish at a point where $\theta=0$, i.e. $V_{aa}-\epsilon_0>0$, and there will be no contribution to Eq.\,(\ref{eq:Stokes}) in this case. 

The theory of complex functions indicates that the zeroes of $V_{ab}$ come by pairs, with an opposite circulation of the phase $\phi$ around the two zeroes of the pair. If $V_{aa}-\epsilon_0$ keeps the same sign at the location of the two zeroes, the singular contributions to Eq.\,(\ref{eq:Stokes}) of the two members of the pair thus compensate each other. To obtain a non-zero flux through the unit cell of the OFL, there must be at least one pair of zeroes of $V_{ab}$ that is "rectified", i.e., for which $V_{aa}-\epsilon_0$ changes sign (hence $\theta$ switches from $0$ to $\pi$) between the two pair members.

Consider now the coupling $\hat V(\bs r)$ given in Eq.\,(\ref{eq:V_DS}). The off-diagonal matrix element $V_{ab}(\bs r)=V_0\,\alpha_+^*(\bs r)\alpha_-(\bs r) $ vanishes at the zeroes of the two coefficients $\alpha_\pm$. If the zeroes of $\alpha_+$ and $\alpha_-$ do not overlap, then the rectification described above cannot operate. All zeroes of $\alpha_+$ lead to a positive sign of $V_{aa}-\epsilon_0=V_0(|\alpha_-|^2-|\alpha_+|^2)/2$ and their contributions to Eq.\,(\ref{eq:Stokes}) thus compensate. Similarly all zeroes of $\alpha_-$ lead to a negative sign of  $V_{aa}-\epsilon_0$, hence no contribution to Eq.\,(\ref{eq:Stokes}) either: one cannot reach a non-zero flux in this case. The only way to circumvent this conclusion is to allow at least some partial overlap between the zeroes of $\alpha_+$ and $\alpha_-$. However in this case, $\hat V$ vanishes at these common zeroes, and the two eigenstates of $\hat V$ coincide. As we show in the main text, this local degeneracy does not prevent the achievement of a series of topological bands.

\vskip 5mm

\centerline{\textbf{S2. Ideal Chern band character}}
\vskip 2mm

Recent theoretical developements introduced a class of `ideal' Chern bands, which share the same algebraic structure as Landau levels despite a non-uniform Berry curvature \cite{roy_band_2014,wang_exact_2021,ledwith_vortexability_2023,estienne_ideal_2023}. These bands are characterized by a specific structure of the quantum geometrical tensor $[\mathcal{Q}_{n,\qbold}]$, defined  by 
\[
[\mathcal{Q}_{n,\qbold}]_{ij}=\bra{\partial_{q_i}\psi_{n,\qbold}}(1-\ket{\psi_{n,\qbold}}\bra{\psi_{n,\qbold}})\ket{\partial_{q_j}\psi_{n,\qbold}},
\] 
where $\ket{\psi_{n,\qbold}}$ is the Bloch state of  band $n$ and quasi-momentum $\qbold$.

In the case of Landau levels, each band is characterized by a uniform quantum geometrical tensor given by the matrix form
\[
[\mathcal{Q}_{n,\qbold}]=\frac{\ell^2}{2}\begin{pmatrix}
1&-\I\\
\I&1
\end{pmatrix},
\]
where $\ell$ is the magnetic length.
It satisfies the identity $[\mathcal{Q}_{n,\qbold}]\boldsymbol\epsilon_-=\boldsymbol0$ (where $\boldsymbol\epsilon_\pm=(\ebold_x\pm\I\,\ebold_y)/\sqrt2$), which physically means that the dynamics of the momentum component $q_x-\I q_y$ is suppressed. This property is at the heart of the possibility to write Landau levels wavefunctions in terms of holomorphic functions. 

More generally, ideal Chern bands are characterized by the existence of a constant null vector $\wbold$, i.e. $[\mathcal{Q}_{n,\qbold}] \wbold=\boldsymbol 0$ for all $\qbold$, which provides  the same algebraic structure than Landau levels \cite{wang_exact_2021}. 

For the ground band $n=0$ of our OFL, we find that the circular vector $\wbold=\boldsymbol\epsilon_-$ acts as an null vector to a very good approximation (on average over the band, we find $||[\mathcal{Q}_{0,\qbold}]\boldsymbol\epsilon_-||=0.04||[\mathcal{Q}_{0,\qbold}]\boldsymbol\epsilon_+||$ for $V_0=100\hbar\omc$). One then expects a strong analogy between our OFL ground band and the lowest Landau level, from the algebraic description of generic wavefunctions to the physical properties of the expected many-body ground states.


\newpage


\centerline{\textbf{S3. Triangular version of the dark-state lattice}}
\vskip 2mm
\begin{figure}[t]
  \begin{center}
  \includegraphics[scale=0.88,trim=10 15 0 0]{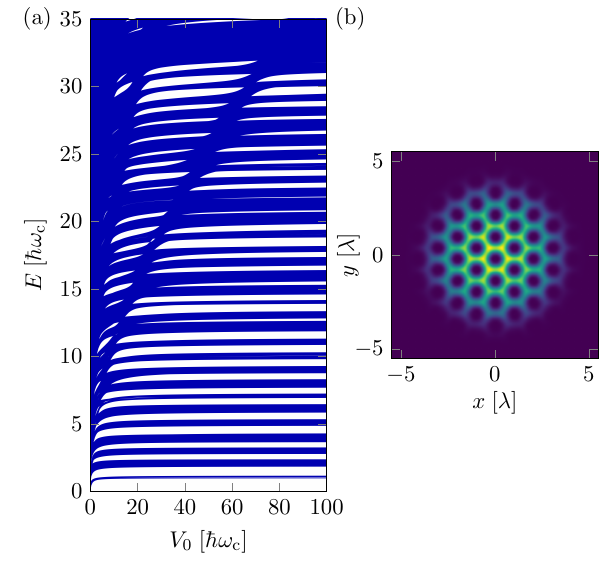}
  \end{center}
  \caption{(a) Bandspectrum of a dark-state OFL with a triangular symmetry, given in Eq.\,(\ref{eq:a_pm_triangle}). We recognize a series of almost equally-spaced bands at low energy, akin to Landau levels. (b) Density profile of a Bose-Einstein condensate in the OFL with triangular symmetry. We used a light shift amplitude  $V_0=100\,\hbar\omc$, an interaction strength $gN/\lambda^2=5\,\hbar\omc$, and a harmonic trap frequency $\omega_{\text{trap}}=0.04\,\omc$.}
  \label{fig:triangle}
\end{figure}

We focused in the main text on a square version of the lattice. We show in this appendix that the same idea can be implemented in a triangular geometry. Consider the lattice formed by superposing three standing waves in the $xy$ plane, superimposed with a running wave propagating along the $z$ axis, so that the amplitudes $\alpha_\pm$ entering in the coupling matrix $\hat V$ of Eq.\,(\ref{eq:V_DS}) read:
\begin{equation}
\begin{array}{rcl}
\alpha_+ &=& c_1+c_2+c_3+\frac{3}{2} \\
\alpha_- &=& c_1-\frac{1}{2}\left(c_2+c_3\right)\ + \ \I\frac{\sqrt 3}{2}\left(c_2-c_3\right)
\end{array}
\label{eq:a_pm_triangle}
\end{equation}
where $c_i=\cos(\bs k_i\cdot\bs r)$ and the wave vectors $\bs k_i$ are at 120° from each other. The zeros of $\alpha_-$ are found at two sets of points. The first one corresponds to the locations where all $c_j=1$, the second one to the locations where all $c_j=-1/2$. The second set coincides with the locations of the zeroes of $\alpha_+$, so that the coupling matrix $\hat V$ vanishes at these points. We recall that this local degeneracy is required for a dark-state based OFL (see appendix S1). 

We show in \fig{fig:triangle}a the energy spectrum as a function of the light coupling strength $V_0$. For  large enough $V_0$, we find a series of  almost equally-spaced energy bands, with a spacing $\hbar\omc=\sqrt{3}\hbar^2 k^2/(2\pi m)$ consistent with the cyclotron frequency expected for the magnetic flux density. For $V_0=100\,\hbar\omc$,  we find a topological lowest band with a flatness ratio $\simeq 5$. An example of Bose-Einstein condensate with a triangular vortex lattice in this configuration is shown in \fig{fig:triangle}b.

\vskip 10mm

\centerline{\textbf{S4. Topological robustness of the ground band}}
\vskip 2mm
\begin{figure}[t]
  \begin{center}
  \includegraphics[scale=0.82,trim=14 15 0 0]{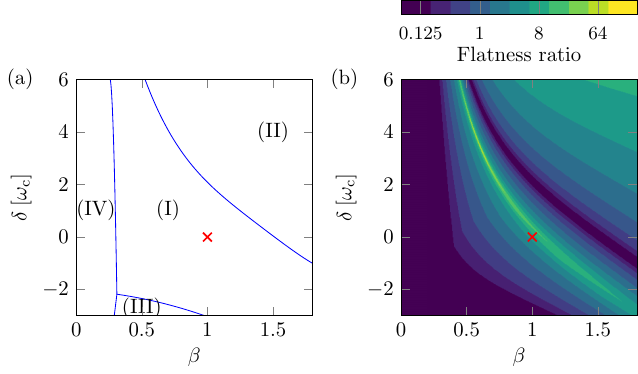}
  \end{center}
  \caption{(a) Phase diagram for the ground energy band of the OFL as a function of the parameters $\beta$ and $h_z$ for $V_0=100\,\hbar\omc$. The phase (I) exhibits a gapped ground band with a Chern number $\mathcal{C}=1$. The phases (II) and (III) are both gapped with a Chern number $\mathcal{C}=0$. In the phase (IV), the ground band overlaps with excited Bloch bands. The blue lines indicate the gapless transition lines separating the different phases. (b) Evolution of the flatness ratio with $\beta$ and $\delta$. The red cross corresponds to the values $\beta=1$ and $\delta=0$ considered in the main text. A maximum flatness ratio of about 130 is obtained for $\beta=0.65$ and $\delta=2.9\,\omega_{\text{c}}$ (barely visible on the plot due the strong sensitivity of high flatness ratios to $\beta$ and $\delta$).}
  \label{fig:flatness}
\end{figure}

The lattice considered in the main text, described by Eqs.\,(\ref{eq:V_DS}-\ref{eq:am2}), depends only on the parameter $V_0$. Moreover the low-energy band structure is essentially independent of $V_0$ for large values of this parameter (see Fig.\,\ref{fig:bands}). We consider here the sensitivity of our results to different kinds of imperfections, namely a modification of the light field profile and an external  magnetic field. More specifically, we consider light field amplitudes $\alpha_+$ and $\alpha_-$ given by
\begin{eqnarray}
\alpha_+(\bs r)&=&\sin X+{\rm i}\sin Y,
\label{eq:am1_bis}
\\
\alpha_-(\bs r)&=& \beta\left(\cos X+\cos Y\right),
\label{eq:am_beta}
\end{eqnarray}
parametrized by $\beta$. We also study the influence of a magnetic field along $z$ giving rise to a Zeeman splitting $\hbar\delta \hat\sigma_z/2$. Equivalently, such a splitting also arises from a detuning $\delta$ with respect to the Raman resonance. The band structure presented in the main text correspond to the values $\beta=1$ and $\delta=0$.

We first consider the phase diagram of the system, more specifically the topological nature of the ground band. As shown in \fig{fig:flatness}a, we find a gapped topological band with Chern number $\mathcal{C}=1$ for a wide range of parameters around $(\beta,\delta)=(1,0)$ (phase I). When increasing the magnitude of the Zeeman field, one finds transition lines at which the gap to the first excited band vanishes, marking the transition to non-topological bands (phases II and III). For weak values of $\beta$, the system becomes metallic (phase IV).

We then consider the flatness ratio of the lowest band, i.e., the ratio of the gap to the first excited band and the width of this lowest band (see Fig.\,\ref{fig:flatness}b). Fine tuning of the parameters $\beta$ and $\delta$ allows to flatten the ground band significantly, with flatness ratios exceeding 100.  This fine tuning should ease the reach of fractional quantum Hall states, for which the interaction energy dominates over the kinetic energy (i.e. the band width).


\clearpage

\end{document}